\documentclass[journal,transmag]{IEEEtran}
\usepackage[top=0.56in, bottom=0.56in, left=0.56in, right=0.56in]{geometry}
\IEEEoverridecommandlockouts
\usepackage{booktabs}
\usepackage{setspace}
\usepackage{multirow}
\usepackage{color}
\usepackage{float}
\usepackage{cite}
\usepackage{enumitem}  
\usepackage[utf8]{inputenc}
\usepackage{lmodern}
\usepackage{tabularx}
\usepackage{amssymb}
\usepackage{csquotes}
\usepackage[justification=centering]{caption}
\usepackage{blindtext}
\usepackage{wrapfig}
\usepackage{graphicx}
\usepackage[bottom]{footmisc}
\DeclareGraphicsExtensions{.eps,.pdf,.gif,.png,.jpg}
\usepackage{amsthm}
\usepackage{array}
\usepackage{algorithm,algorithmic}
\usepackage{booktabs}
\usepackage{graphicx}
\usepackage{epstopdf}\usepackage[utf8]{inputenc}
\usepackage[hyphens,spaces,obeyspaces]{url}
\usepackage[english]{babel}
\usepackage{verbatim} 
\usepackage{lineno,nohyperref}
\usepackage{epstopdf} 
\modulolinenumbers[5]

\RequirePackage{filecontents}
\usepackage{ragged2e}
\usepackage{url} 
\usepackage[T1]{fontenc}
\usepackage[utf8]{inputenc}
\usepackage{microtype} 

\newcommand{\vect}[1]{\boldsymbol{#1}}
\newcommand*{\email}[1]{#1}
\PassOptionsToPackage{hyphens}{url}
\newtheoremstyle{mystyle}
{}
{}
{\itshape}
{}
{\bfseries}
{.}
{ }
{\thmname{#1}\thmnumber{ #2}\thmnote{ (#3)}}
\newcommand*{\defeq}{\mathrel{\vcenter{\baselineskip0.5ex \lineskiplimit0pt
			\hbox{\scriptsize.}\hbox{\scriptsize.}}}%
	=}
\theoremstyle{mystyle}

\newcounter{subassumption}[asu]

\makeatletter
\renewcommand{\p@subassumption}{\theasu}
\makeatother
\usepackage{amsthm}
\usepackage{xpatch,amsthm}
\makeatletter
\xpatchcmd{\@thm}{\fontseries\mddefault\upshape}{}{}{} 
\makeatother

\makeatother
\usepackage{cite}
\usepackage{amsmath,amssymb,amsfonts}
\usepackage{algorithmic}
\usepackage{graphicx}
\usepackage{textcomp}
\usepackage{xcolor}
\def\BibTeX{{\rm B\kern-.05em{\sc i\kern-.025em b}\kern-.08em
		T\kern-.1667em\lower.7ex\hbox{E}\kern-.125emX}}

\begin{document}
	\title{Digital Twin Backed Closed-Loops for Energy-Aware and Open RAN-based Fixed Wireless Access  Serving Rural Areas}
	 \author{Anselme~Ndikumana,~\IEEEmembership{Member,~IEEE,~}
		Kim~Khoa~Nguyen,~\IEEEmembership{Senior~Member,~IEEE,~}\\
		and~Mohamed~Cheriet,~\IEEEmembership{Senior~Member,~IEEE,~}
		\IEEEcompsocitemizethanks{
			\IEEEcompsocthanksitem Anselme Ndikumana, Kim Khoa Nguyen, and Mohamed Cheriet are  with Synchromedia Lab, École de
			Technologie Supérieure, Université du Québec, QC, Canada, E-mail: (\email{anselme.ndikumana.1@ens.etsmtl.ca; kim-khoa.nguyen@etsmtl.ca; Mohamed.Cheriet@etsmtl.ca}).
		\IEEEcompsocthanksitem 	``This work was supported by NSERC (under project ALLRP
		566589-21) and InnovÉÉ (INNOV-R program).''.
		}}
	\maketitle
	\begin{abstract}

Internet access in rural areas should be improved to support digital inclusion and 5G services. Due to the high deployment costs of fiber optics in these areas, Fixed Wireless Access (FWA) has become a preferable alternative. Additionally, the Open Radio Access Network (O-RAN) can facilitate the interoperability of FWA elements, allowing some FWA functions to be deployed at the edge cloud. However, deploying edge clouds in rural areas can increase network and energy costs. To address these challenges, we propose a closed-loop system assisted by a Digital Twin (DT) to automate energy-aware O-RAN based FWA resource management in rural areas. We consider the FWA and edge cloud as the Physical Twin (PT) and design a closed-loop that distributes radio resources to edge cloud instances for scheduling. We develop another closed-loop for intra-slice resource allocation to houses. We design an energy model that integrates radio resource allocation and formulate ultra-small and small-timescale optimizations for the PT to maximize slice requirement satisfaction while minimizing energy costs. We then design a reinforcement learning approach and successive convex approximation to address the formulated problems. We present a DT that replicates the PT by incorporating solution experiences into future states. The results show that our approach efficiently uses radio and energy resources.
\end{abstract}

\section{Introduction}
\label{sec:introduction}
Today's world relies on Internet connectivity more than ever. However, some parts of the world remain unconnected, especially in rural areas where deploying fiber optics is not economically viable \cite{adityo20215g}. 5G Fixed Wireless Access (5G FWA) could be a solution for rapid deployment and reducing network costs in these areas. 5G FWA enables operators to provide wireless connectivity through radio links to areas that lack the infrastructure for wired Internet solutions \cite{laraqui2017fixed}. In the 5G FWA setup, houses are equipped with Customer Premises Equipment (CPE), where a 5G wireless air interface connects the CPE to a radio unit.  5G FWA is the fastest-growing broadband segment, boasting a 71\% compound annual growth rate. It is estimated that 5G FWA will surpass 58 million subscribers by 2026 \cite{5gamericas}.

Low-cost and high-automation technologies such as Network Function Virtualization (NFV), Virtual Network Function (VNF), and Open Radio Access Network (O-RAN) \cite{allianceORANUseCases} can help to improve 5G FWA. In such an O-RAN based 5G FWA network, each house is connected to the O-RAN Radio Unit (O-RU) by a CPE. Each O-RU is linked to the O-RAN Distributed Unit (O-DU) hosted as VNF at the edge cloud. In rural areas, this edge cloud can be powered by renewable energy, supplemented by the power grid. Consequently, a joint radio and energy resource management framework is essential to manage the system effectively.
	
We can use a Digital Twin (DT) to represent an O-RAN based FWA, where the DT is a reflection of the O-RAN based FWA processes and states. Since the DT is a virtual representation of the Physical Twin (PT) \cite{isik2023architectural} \cite{mihai2022digital}, in our case, the PT is the O-RAN based FWA serving rural areas. The DT continuously communicates with the PT to collect environmental states, predictively optimizing radio resource allocation and energy consumption. Like simulations, the DT virtually represents the physical object; however, their scales differ. A simulation is typically a single process and therefore cannot fully represent a physical object. In contrast, a DT \cite{haag2018digital} runs multiple simulations to examine various aspects and functions of a physical object (i.e., the PT). The key benefit is that the DT of an O-RAN based FWA in one rural area can help replicate the O-RAN based FWA in other rural areas, including elements, processes, firmware, and dynamics of the PT.

In O-RAN based FWA serving rural areas, we assume the end-users of CPEs require enhanced mobile broadband (eMBB) services with different Quality of Service (QoS), such as mixed reality, video streaming, home gaming, remote learning, and in-home healthcare. For example, the network traffic of home gaming and in-home healthcare \cite{martiradonna2021cascaded} services should be treated differently because in-home healthcare is much more critical. Therefore, to accommodate these diverse QoS needs, network slicing \cite{chuah2020intelligent} that treats network traffic differently during the  resource allocation should be considered for the O-RAN based FWA. However, because of the dynamic nature of network slicing operations, human-in-the-loop service orchestration management is challenging when trying to meet the demands of network slices \cite{boutaba2021ai}. Therefore, zero-touch network automation supported by DT is substantial for successfully managing network resources in rural areas.

In O-RAN based FWA resource management, a small timescale for energy scheduling, such as an hour, is large compared to radio resource scheduling, which operates on the order of seconds or milliseconds. Therefore, jointly optimizing radio resource allocation and energy costs in O-RAN based FWA presents significant challenges. This paper develops two closed loops for energy-efficient radio resource management in O-RAN based FWA, assisted by DT technology, to address these challenges. The main contributions of this work are as follows:
\begin{itemize} 
	\item
	In PT, at Near-RT RIC, we develop one closed-loop to distribute radio resources to  O-DUs for scheduling. We propose another closed-loop for intra-slice operations at O-DU to allocate radio resources to CPEs. Both closed-loops monitor radio resource utilization and extract knowledge from radio resource utilization data to perform zero-touch Resource Block (RB) adjustments to meet slice requirements. Existing literature focus on FWA  \cite{rahmawati2022assessing} \cite{lappalainen2021planning}. So far, the literature has never discussed zero-touch and data-driven multi-level closed-loops for radio resource scheduling in O-RAN based FWA.	
	\item
	We propose an energy model for PT powered by the grid and renewable energy sources to serve the edge cloud that hosts O-RAN elements and allocates RBs to CPEs. Then, we join the energy model with the communication model to maximize radio resource utilization and slice requirement satisfaction while minimizing energy costs. This joint problem has not yet been tackled in the literature.
	\item
	Considering O-RAN based FWA radio and energy resources modeling, where a small timescale for the energy is large for the communication network, we formulate ultra-small time and small timescale optimization problems for PT to minimize energy cost while maximizing radio resource utilization and meeting slice requirements. Then,  we design reinforcement learning and Successive Convex Approximation (SCA) approaches to handle the formulated non-convex problems.   
	\item  We use energy and communication models' solution data to construct the DT for predicatively optimizing radio resource allocation and energy consumption. We continuously feed PT data into the DT to predict the required radio and energy resources. In other words, the DT mirrors the most recent representative characteristics of the PT, where data and Continual Learning (CL) \cite{van2019three} establish the link between the PT and its digital counterpart, the DT. There is a lack of datasets for O-RAN based FWA in rural areas in the literature. Utilizing solution data from PT can help us overcome this challenge. To the best of our knowledge, leveraging O-RAN based FWA data and CL for radio and energy resource management in a DT environment has not yet been addressed in the literature.
\end{itemize}

The rest of the paper is structured as follows. Section \ref{sec:LiteratureReview}  discusses the related works, and Section  \ref{sec:system-model} presents the
system model. Section \ref{sec:EnergyEfficient} discusses energy-efficient radio resource allocation, and Section \ref{sec:ProblemFormulation} presents the problem formulation. Section \ref{sec:ProposedSolution} discusses the solution approach. Section \ref{sec:PerformanceEvaluation} presents a performance evaluation, while  Section \ref{sec:Conclusion} concludes the paper.

\section{Literature Review}
\label{sec:LiteratureReview}
The related works are organized into three categories: $(i)$ FWA and network slicing, $(ii)$ energy-efficient networks and resource allocation, and $(iii)$ closed-loops and DT in resource allocation.

\emph{FWA and network slicing:}
The authors in \cite{hashemi2017integrated} discussed integrating access and backhaul networks in FWA to extend coverage in serving residential users in suburban areas. In \cite{rahmawati2022assessing}, the authors considered urban residential areas as a case study for FWA. 
They introduced a novel method to calculate the necessary number of base stations for capacity and coverage planning. In their work \cite{lappalainen2021planning}, the authors proposed a new approach for FWA to find the maximum number of households that can achieve target minimum bit rates simultaneously on the uplink and downlink, given network resources and a cell radius. Their approach specifically discussed rural areas utilizing Multiple-Input Multiple-Output (MIMO) FWA technology to deliver fixed broadband services to households. Regarding network slicing in FWA, the authors in \cite{matrakidis2021converged} used a testbed to exploit QoS-aware end-to-end slicing. Compared to these works, our approach considering O-RAN based FWA  PT and DT is new in the literature.
\begin{figure*}[t]
	\centering
	\includegraphics[width=2.0\columnwidth]{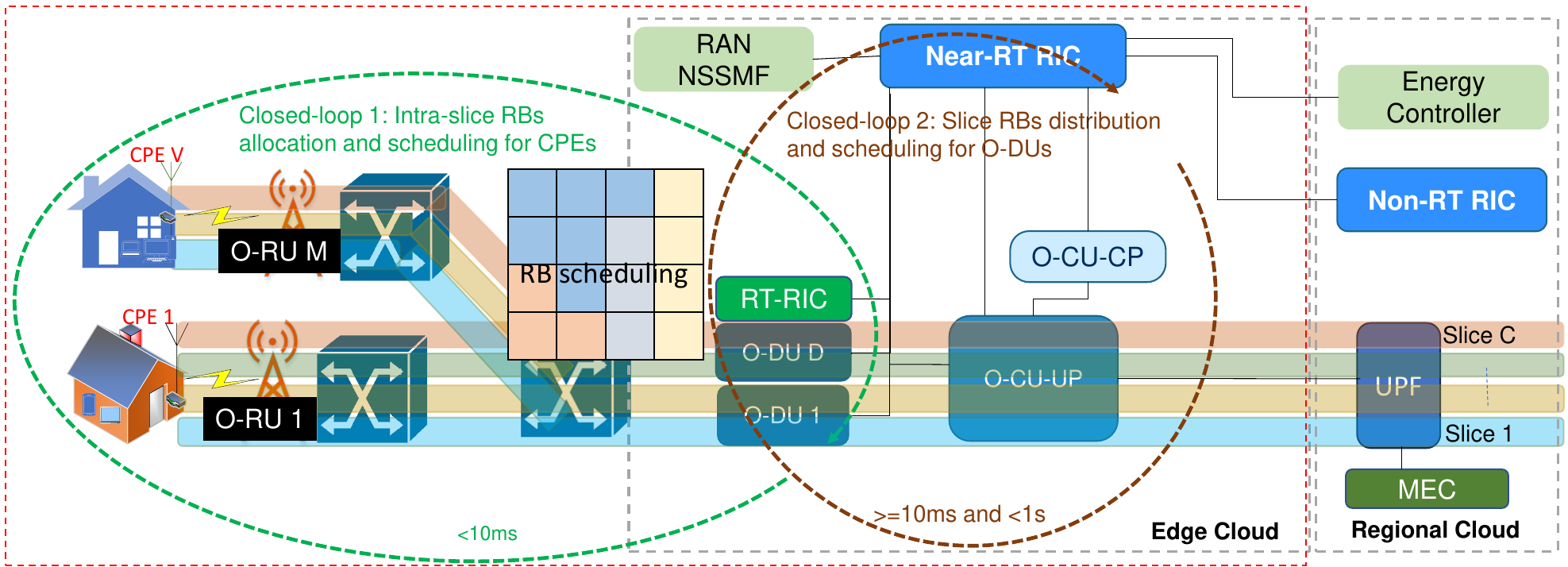}
	\caption{System model for PT (in dashed red rectangle).}
	\label{fig:SystemModel}
\end{figure*}

\emph{Energy-efficient networks and resource allocation:} The authors in \cite{larsen2021energy} introduced an energy consumption modeling approach to minimize energy usage in the Cloud RAN. The authors in \cite{kaliski2023supporting} targeted cost-effective and energy-efficient network deployment through designing an O-RAN architecture incorporating Multi-access Edge Computing (MEC) and multi-cell connectivity. In \cite{miyazawa2024energy}, the authors controlled power-on/off states of O-RAN base stations according to the predictions of a pedestrian flow. When O-RAN base stations are turned off due to low pedestrian flow, users can access a non-terrestrial network, thereby reducing power consumption. The authors in \cite{dinh2020home} considered an energy model that includes renewable energy, energy storage, and energy from the power grid for home usage. However, their proposed energy model does not consider the communication model. In \cite{azimi2021energy}, the authors proposed an energy-efficient resource allocation approach for the RAN that incorporates network slicing. Furthermore, the authors in \cite{pamuklu2021energy} optimized radio resource allocation and DU selection in O-RAN, aimed at reducing energy consumption.  However, none of these prior studies has ever addressed multi-level resource allocation, which involves interactions between multiple closed-loops.

\emph{Closed-loops and  DT in resource allocation:}
The authors in \cite{boutaba2021ai} discussed artificial intelligence-based closed-loop automation and examined the challenges associated with closed-loop automation in 5G networks. In \cite{chang2018radio}, the authors explored RAN inter-slice resource partitioning and allocation, formulating an optimization problem for resource allocation that enables inter-slice radio resource sharing. Additionally, in \cite{xie2019towards, naik2022closed}, authors proposed a closed-loop-based solution for automatic slicing assurance in 5G RAN. Furthermore, in \cite{9833928}, a CL framework was proposed to model the evolving affinity between PT and Cyber Twin (CT), enabling synchronization between them. In \cite{peng2022distributed}, the authors discussed interactions among DTs using credit-based incentives to determine the resource allocation strategy for mobile devices offloading data to edge clouds. However, the joint problem of energy and radio resource allocation for a 5G FWA network in the edge cloud using more than one closed-loop and DT has not yet been tackled in the literature. 

In our prior work \cite{10437288}, we presented an energy controller for optimizing energy consumption, a Multi-Access Edge Computing (MEC) server that hosts DT, and a User Plane Function (UPF) for data processing at the edge cloud. However, since rural areas have lower population densities compared to urban and suburban areas,  radio resources are often underutilized due to low network traffic. Therefore, in this work, we consider a different network architecture in which the energy controller, UPF, and MEC server are all located in the regional cloud separately from the PT that has the edge cloud. Also, in many rural areas with limited access to the power grid, solar energy can effectively complement grid energy. The edge cloud can be connected to the grid and solar panels. During sunny days, the solar panels generate electricity, reducing reliance on the grid. Our approach offers several advancements over previous methods. Firstly, we employ two-level closed-loops that exchange information to enhance radio resource allocation and reduce energy costs in O-RAN based FWA, considering both grid and solar energy. Secondly, our approach involves a DT that replicates PT behavior by integrating solution experiences into future states. Lastly, to the best of our knowledge, this is the first work that integrates energy cost minimization and radio resource allocation using two-level closed-loops and DT in O-RAN based FWA serving rural areas.
\begin{table}[t]
	\caption{Summary of key notations.}
	\label{tab:table1}
	\begin{tabular}{ll}
		\toprule
		Notation & Definition\\
		\midrule
		$\mathcal{V}$ & Set of CPEs, $|\mathcal{V}|= V$\\
		$\mathcal{M}$ & Set of O-RUs,  $|\mathcal{M}|= M$\\
		$\mathcal{P}$ & Set of Server, $|\mathcal{P}|= P$\\
		$\mathcal{D}$ & Set of O-DUs, $|\mathcal{D}|= D$\\
		$\mathcal{E}$ & Set of O-CUs, $|\mathcal{E}|= E$\\
		$\mathcal{K}$ & Set of services, $|\mathcal{K}|= K$\\
		$\mathcal{\beta}$ & Set of RBs  $|\mathcal{\beta}|=\beta$\\
		$\mathcal{C}$ & Set of slices $|\mathcal{C}|=C$\\
		$\lambda_{v}^{k,p}$ & Arrival rate  for service $k$\\
		$\Psi_d^{c,k}$ & Queue status parameter\\
		$\mu_{v}^{k,p}$ & Service rate for service $k$\\
		$\Gamma^k$ & Delay budget for service $k$\\
		$R^v_m$ & Data rate of each CPE $v$ \\
		$\varphi^{c,k}$ & Slice requirement satisfaction\\
		$\tilde{\varphi}^{c,k}_d$ & RB usage ratio at O-DU $d$ \\
		$\Omega_d^{c,k}$& Intra-slice orchestration parameter \\
		$\chi_{m,v}$& Distance between CPE $v$ and O-RU $m$\\ 
		$L(t)$ &  Total available energy at time $t$ \\
		$L_{cons}(t)$ & Energy consumption at time $t$ \\
		$\pi(P)$ & Power consumption of the servers $P$\\
		$H(t)$ & Energy cost for edge cloud\\
		\bottomrule
	\end{tabular}
\end{table}
\section{System model for Physical Twin}
\label{sec:system-model}
We present PT in Fig \ref{fig:SystemModel}, while DT will be presented in Section \ref{sec:ProposedSolution}. We summarize the notations used in this paper in Table \ref{tab:table1}.

\begin{figure}[t]
	\centering
	\includegraphics[width=1.0\columnwidth]{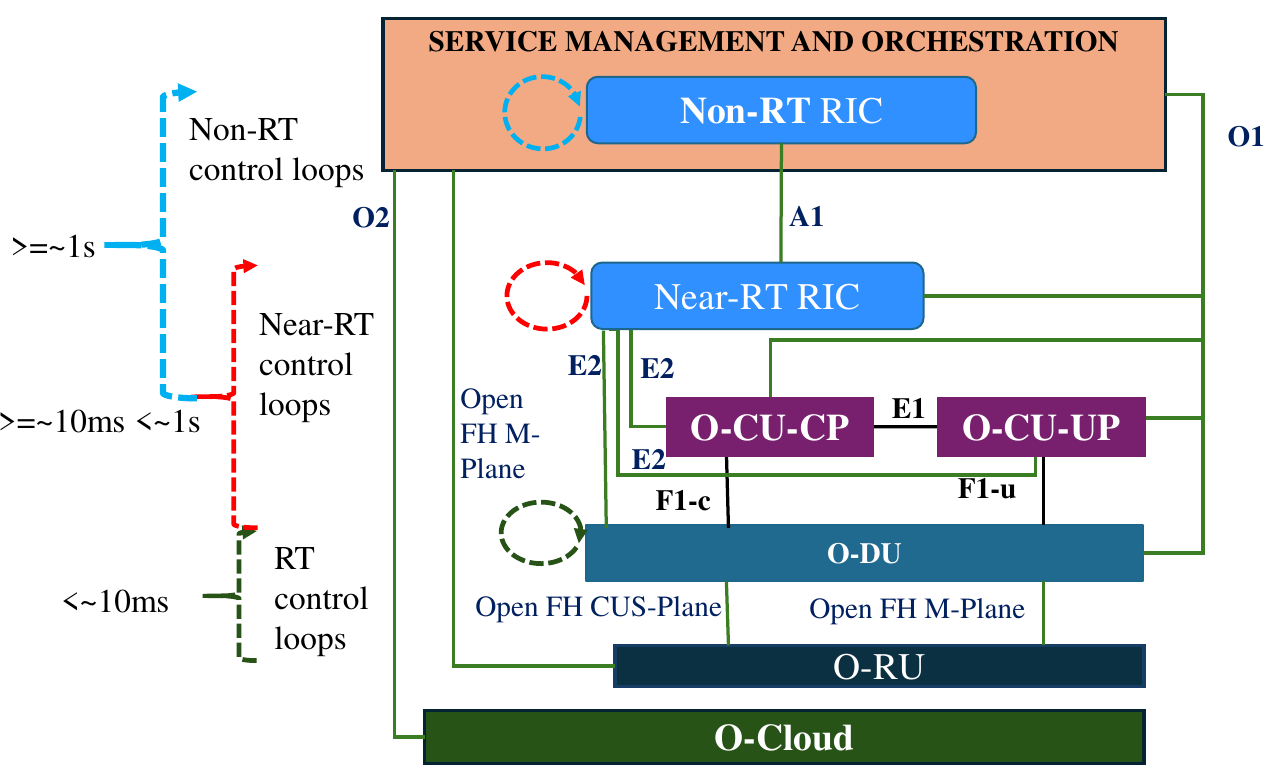}
	\caption{O-RAN and its control loops \cite{allianceORANUseCases}.}
	\label{fig:O-RAN_Control_loops}
\end{figure}

In Fig. \ref{fig:SystemModel}, we denote the set of houses as $\mathcal{V} = \{ 1, \dots,V \}$, where each house $v \in \mathcal{V}$ is equipped with CPE comprising rooftop antenna for Internet access. Unless specified otherwise, we use the terms CPE and house interchangeably. Similar to \cite{saba2024using}, CPEs are wirelessly connected to O-RUs via line-of-sight (LoS) paths. We define $\mathcal{M} = \{ 1, \dots,M \}$ as the set of O-RUs. Each O-RU $ m \in \mathcal{M} $ is connected to an O-DU via a wired fronthaul network, where the O-DU performs RB scheduling for the CPEs. We consider multiple O-DUs, each serving more than one O-RU, and denote the set of O-DUs as $\mathcal{D} = \{ 1, \dots,D\}$.
O-DUs are connected to the O-RAN Central Unit Control Plane (O-CU-CP) and the O-RAN Central Unit User Plane (O-CU-UP). Furthermore, our model incorporates O-RAN closed-loops and distributed intelligent controllers like the Near-Real-Time RAN Intelligent Controller (Near-RT RIC) and the Non-Real-Time RAN Intelligent Controller (Non-RT RIC).  We utilize two types of control-loops (loops $1$ and $2$) of O-RAN shown in Fig. \ref{fig:O-RAN_Control_loops}. Loop $1$ operates at a timescale of less than $10$ milliseconds and can be used for RB scheduling at the O-DU. Loop $2$ operates within the Near-RT RIC on a timescale greater than or equal to $10$ and less than $1000$ milliseconds, making it suitable for radio resource optimization. In our model, the Near-RT RIC, O-DU, O-CU-UP, and O-CU-CP are implemented on servers at the edge cloud, where $\mathcal{P}$ is a set of servers. For loop $2$, we assume that the delay between the O-DU,  Near-RT RIC, O-CU-UP, and O-CU-CP is negligible because these O-RAN elements are co-located within the same edge cloud. Furthermore, we consider a regional cloud that hosts the Non-RT RIC and the MEC server outside the PT. The MEC server, accessible via UPF \cite{ndikumana2019joint, ndikumana2022age}, is used to build the DT. In Fig. 1, the grey dotted boxes indicate the boundaries of the edge and regional clouds, while the red dotted box indicates the boundary of the PT.
\begin{figure}
	\centering
	\includegraphics[width=0.95\columnwidth]{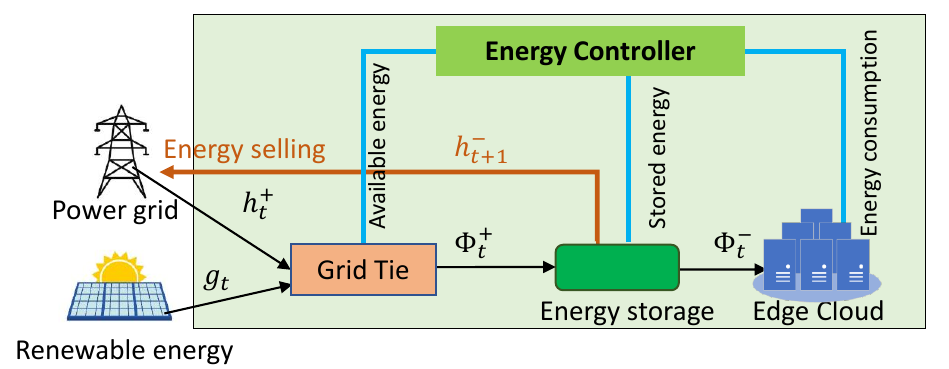}
	\caption{Energy model for PT.}
	\label{fig:EnergyModel}
\end{figure} 

We assume that the RBs $\mathcal{\beta}$ and the computation resources available on edge cloud servers can be divided and allocated to network slices that serve CPEs. Let $\mathcal{K}$ denote the set of services required by CPE users, with $\Gamma^k$ representing the delay budget for each service $k$. Network slicing is employed to meet diverse QoS requirements, such as latency, where each service is assigned to a specific slice. The set of slices is represented by $\mathcal{C} = \{ 1, \dots,C\}$. The Near-RT RIC receives service profiles from service providers via the RAN Network Slice Subnet Management Function (NSSMF). It then performs the initial mapping of services to slices and the mapping of slices to O-RAN elements, such as O-DUs. The Near-RT RIC also handles the initial distribution of RBs to the slices associated with O-DUs using a closed-loop approach (loop 2) for scheduling. Each O-DU can manage multiple slices. Subsequently, within closed-loop 1, each slice at the O-DU allocates RBs to its associated CPEs. Closed-loop 1 uses a Real-Time RAN Intelligent Controller (RT RIC) for saving and sharing information related to slices with closed-loop $2$ so that the closed-loop $2$ can adjust RB allocated to O-DUs. Then, closed-loop $2$ sends updated RBs to closed-loop $1$ so that the closed-loop $1$ can adjust intra-slice RB allocation. Furthermore, based on the utilization ratio of RB and the expected number of packet arrivals defined in the next section, an O-DU or a server that does not serve any CPE can be disabled to minimize the energy consumption of the edge cloud. In other words, O-RAN elements that do not have active slices can be disabled.

The edge cloud hosting O-RAN elements requires energy to operate. As shown in Fig. \ref{fig:EnergyModel}, we consider two sources of energy reaching the edge cloud: the power grid and renewable energy. The energy passes through a grid-tie device, which combines energy from the power grid and renewable energy \cite{deng2013multigreen}. Also, there is an energy storage that stores renewable energy when its generation is high or grid energy when the price of the energy grid is very low. Energy storage provides energy to edge cloud, and surplus energy can be sold to the energy market.

\section{ Energy Efficient Radio  Resource Management for Physical Twin}
\label{sec:EnergyEfficient}
This section introduces the communication model for 5G FWA and the energy model for the edge cloud.

\subsection{Communication Model of 5G FWA}
\label{subsec:CommunicationModel}
\subsubsection{Initial RB Distribution to O-DU}
\label{subsubsec:RBsDistributionO-DUs}

The Near-RT RIC initially employs a round-robin policy \cite{kanhere2002fair}, cyclically assigning slices associated with services starting from O-DU 1 until all slices are created or all O-DUs are utilized. However, the round-robin policy is not restrictive; other policies can also be applied. Then, at Near-RT RIC, the closed-loop $2$ distributes available RBs $ \mathcal{\beta}$ to O-DUs  such that $\beta=\sum_{d=1}^{D}\beta_d$. We denote  $\beta_d=\lfloor\frac{\beta}{D}\rfloor $ as RB distributed to each O-DU $d$ for scheduling. The RBs $\beta_d$ available at each O-DU $d$ are allocated to the slices using that O-DU. Furthermore, we consider CPEs that use RBs $ \mathcal{\beta}$ managed by slices $C$ at O-DUs. Based on RB utilization and slice requirement satisfaction, the closed-loop $2$ can dynamically update RBs distributed to O-DUs using the zero-touch RB management approach that will be discussed in this section.

The closed-loop $2$ allocated RB $\beta^{c,k}_d$ to each slice $c$ of service $k$ at O-DU $d$ such that: 
 \begin{equation}
 	\sum_{k=1}^{K_d}x^{c,k}_{d} \beta^{c,k}_d\leq \beta_d,
 \end{equation}	
 where  $K_d$ represents the number of services that utilize  O-DU $d$. We define  $x^{c,k}_{d}$ as RB allocation decision variable, where $x^{c,k}_{d}$ is expressed as follows:
 \begin{equation}
 	\label{eq:mappingslice_v0du}
 	\setlength{\jot}{10pt}
 	x^{c,k}_{d}=
 	\begin{cases}
 		1,\; \text{if $\beta^{c,k}_d$ is allocated to slice $c$}\\
 		       \text{ of service $k$ at O-DU $d$,}\\
 		0, \;\text{otherwise.}
 	\end{cases}
 \end{equation}
 We consider initial RB $\beta^{c,k}_d$ can be obtained via  auction mechanism discussed in our previous work  in \cite{ndikumana2022two}. Furthermore, we impose the following constraint to ensure that one O-DU serves slice $c$ of service $k$:
 \begin{equation}
 	\begin{aligned}
 		\sum_{c\in \mathcal{C}}x^{c,k}_{d} \leq 1 , \;  \forall d, k.	\end{aligned}
 \end{equation}

\subsubsection{Initial Intra-slice RB Allocation for CPEs}
\label{subsubsec:Intra-slicesRBs}
Inside each slice $c$, we use closed-loop 1 to assign RB to CPEs connected to O-RUs, where 
we assume that each O-RU serves the CPEs within its coverage area. Based on the selected 5G numerology $i$ defined in \cite{138211}, each RB $\beta^{c,k}_d$ is divided into
 $f^{c,k}_{i,d}$ number of sub-bands, indexed by $\mathcal{F}^{c,k}_{i,d} = \{1, 2, \dots ,F^{c,k}_{i,d}\}$ in the frequency-domain and $t^{c,k}_{i,d}$ number of transmission time intervals, indexed by $\mathcal{T}^{c,k}_{i,d} = \{1, 2, \dots , T^{c,k}_{i,d}\}$ in the time-domain. We remind that each service is associated with one slice, where more than one CPE can request one service.

Based on data $R^{c,k}_v$ from upper layer reaching slice $c$ of service $k$ at O-DU $d$ that needs to be sent to CPE $v$, we can calculate the required RB $\beta_v$ for CPE $v$ as follows: 
\begin{align}
	\label{Rb_2}
	\beta_v =
	\frac{10^{6}\cdot R^{c,k}_v \cdot t^{i}}{ \sum_{j=1}^{J}(\vartheta^{(j)}_{layers} \cdot Q^{(j)}_{mcs}  \cdot \zeta^{(j)} \cdot 12 \cdot R_{max}\cdot(1-{OH}^{(j)})}.
\end{align} 
$J$ is the number of component carriers, and $\vartheta^{(j)}_{layers}$  represents the maximum number of MIMO layers. $Q^{(j)}_{mcs}$ denotes the modulation order, and $\zeta^{(j)}$ is a scaling factor. ${OH}^{(j)}$ is used to represent the overhead for the control channel, and $t^{i}$ denotes the average Orthogonal Frequency Division Multiplexing (OFDM) symbol duration in a subframe of numerology $i$. $R_{max}$ represents the maximum coding rate \cite{etsi1308}. 
We define
$y_{v,m}^{\beta}$ as binary decision variable, where $	y_{v,m}^{\beta}$  is given by:
\begin{equation}
	\label{eq:RB_allocation_variable}
	\setlength{\jot}{10pt}
	y_{v,m}^{\beta}=
	\begin{cases}
		1,\; \text{If RB $\beta_v$ is allocated to CPE $v$,}\\
		0, \;\text{otherwise.}
	\end{cases}
\end{equation}
Furthermore, we define the channel coefficient between  CPE $v$ and O-RU on allocated RB
$\beta_v$ as follows:
\begin{equation}
	h_{v}^{t_i,f_i}= \tilde{h}_v^{t_i,f_i}+ e_v^{t_i,f_i},
\end{equation}
where $\tilde{h}_v^{t_i,f_i}$ is the estimated Channel State Information (CSI), and $ e_v^{t_i,f_i}$ is estimated channel error for $f_i \in \mathcal{F}^{c,k}_{i,d}$ and $t_i \in \mathcal{T}^{c,k}_{i,d}$. More details on CSI and channel error estimation are discussed in \cite{yoo2004capacity}.  Then, we  use $h_{v}^{t_i,f_i}$ to compute the following achievable Signal-to-Noise Ratio (SNR) at the CPE $v$ on the RB $\beta_vow$:
\begin{equation}
	\label{eq:decisoF_nariable}
	\delta_v^{t_i,f_i}= 
	\frac{x^{c,d}_{b,k}|h_{v}^{t_i,f_i}|^2 \Lambda_{v, m}}{\sigma_v^2\chi^{\varkappa}_{v,m}},\; 
\end{equation}
where 
$ \Lambda_{v, m}$ denotes the transmission power and $\chi_{m,v}$
is the distance between the CPE $v$ and its O-RU $m$. We use  $\varkappa$  as the path
loss exponent and $\sigma_v^2$ represents the noise power. Since both CPE and O-RU are fixed, we assume that the distance $\chi_{m,v}$ between them does not change, and handover is not required. Therefore, the data rate for the CPE $v$ on the allocated RB $\beta_v$ can be written as:
\begin{equation}
	\label{eq:data_rate}
	\begin{aligned}
		R_{v,m} =y_{v,m}^{\beta}\omega_v^{t_i,f_i}\log_2\left(1 + \delta_v^{t_i,f_i} \right),  \;\forall v \in \mathcal{V},
	\end{aligned}
\end{equation}
where $\omega_v^{t_i,f_i}$ denotes the bandwidth of the RB $\beta_v$. 
We use 3GPP TS 38.306 standard \cite{etsi1308} to approximate the required RBs $\beta_v$ that can achieve data transfer $R^{c,k}_v$. Due to various factors that can affect signal strength and quality such as distance and obstacles, we use  $(8)$ that considers SNR to verify whether data rate $R_{v,m}$ can really support  $R^{c,k}_v$.

\subsubsection{Feedback for Closed-Loops}
\label{subsubsec:Feedbackforclosed}
In previous subsections, we discussed initial RB allocation to O-DUs and CPEs using closed-loops $2$  and $1$. Here,  we discuss the feedback of closed-loops $1$ and $2$ after the initial RB allocation. In other words, the feedback helps to close the two loops.

\emph{Feedback for closed-loop $1$ after initial RB allocation for CPEs:} O-DU monitors RB utilization. Based on incoming data $R^{c,k}_v$ for CPE $v$  that needs service $k$ associated with slice $c$, the queuing delay $q_{v,d}^{c,k}$ is given by:
\begin{equation}
	q_{v,d}^{c,k}= \frac{1}{\lambda_{v}^{k,p} - \mu_{v}^{k,p}},
\end{equation}
where $\lambda_{v}^{k,p}$ is arrival rate and  $\mu_{v}^{k,p}$ is service rate. Subscript $p$ represents the server that hosts O-DU $d$ that serves service $k$.
We use a limited buffer size $\tilde{B}_d^{c,k}$ for service $k$ served by slice $c$ at O-DU $d$. Then, we introduce dynamic queue status parameter $\Psi_d^{c,k}$ and fixed buffer threshold $B_d^{c,k}$, where $\Psi_d^{c,k}$ can  be expressed as follows:
\begin{equation}
	\setlength{\jot}{10pt}
	\Psi_d^{c,k} =\max \{(\tilde{B}_d^{c,k}-{E[\lambda_{v}^{k,p}]}), B_d^{c,k} \},
\end{equation}	
where we denote $E[\lambda_{v}^{k,p}]$ as the expected number of packet arrivals in the queue for service $k$ at server $p$. Furthermore, for transmission and propagation delays, we consider data  $R^{c,k}_v$ of each CPE $v$ using wired fronthaul and wireless networks, where the wireless transmission delay  between CPE and O-RU is given by:
\begin{equation}
	\eta^{c,k}_{v,m}= \frac{R^{c,k}_v}{R_{v,m}}.
\end{equation}
Furthermore, the fronthaul transmission delay $\eta^{c,k}_{m,d}$l between O-RU $m$ and  O-DU $d$ is defined as follows:
\begin{equation}
	\eta^{c,k}_{m,d}=\frac{R^{c,k}_v}{\varpi_{m,d}},
\end{equation}
where $\varpi_{m,d}$ denotes the capacity of fronthaul link between O-RU $m$ and O-DU $d$.  
The fronthaul propagation delay $\eta_{m,d}$ is given by:
\begin{equation}
	\eta_{m,d}=\frac{\rho^{m, d}}{\kappa},
\end{equation}
where $\kappa$ denotes the propagation speed and $\rho^{m, d}$ represents the length of fronthaul link.

We can compute the end-to-end delay as follows:
\begin{equation}
	\eta^{c,k}_{v}= q_{v,d}^{c,k}+ \eta^{c,k}_{v,m} +\eta^{c,k}_{m,d} + \eta_{m,d},
\end{equation}
where we use $\eta^{c,k}_{v}$ as feedback for the closed-loop $1$. Furthermore, we enforce  $\eta^{c,k}_{v}$ to satisfy delay budget by setting  $\eta^{c,k}_{v} \leq \Gamma^k$.  Then,  we define network slice requirement satisfaction $\varphi^{c,k}$ to evaluate whether or not each slice $c$ of  service $k$ meets delay budget $ \Gamma^k$, where $\varphi^{c,k}$ is given by:
\begin{equation}
	\label{eq:network_slice_requirement_satisfaction}
	\varphi^{c,k}=\frac{\sum_{v=1}^{V^k}y_{v,m}^{\beta} \xi_v^{c,k}}{V^k}.
\end{equation}
$V^k$ denotes the number of CPEs that are using service $k$. We define $\xi_v^{c,k}$ as the delay budget satisfaction parameter, where $\xi_v^{c,k}$ is expressed as follows:
\begin{equation}
	\label{eq:satisfaction}
	\setlength{\jot}{10pt}
	\xi_v^{c,k}=
	\begin{cases}
		1,\; \text{if $\eta^{c,k}_{v} \leq  \Gamma^k$}\\
		0, \;\text{otherwise.}
	\end{cases}
\end{equation} 

\emph{Feedback for closed-loop $2$ after initial RB distribution to O-DU:}
We use $\tilde{\varphi}^{c,k}_d$ defined below as RB utilization ratio, where  $\tilde{\varphi}^{c,k}_d$ helps to evaluate the utilization of RB $\beta_d$ allocated to each O-DU $d$ for scheduling:
\begin{equation}
	\label{eq:RBusage}
	\tilde{\varphi}^{c,k}_d=\frac{\sum_{k=1}^{K_d} \beta^{c,k}_d}{\beta_d}.
\end{equation} 
\subsubsection{Zero-Touch RB Management for Closed-loop $1$} 
We discuss slice orchestration parameter $\Omega_d^{c,k}$ and resource management actions for closed-loop $1$ to update initial RB allocation based on network slice requirement satisfaction and RB utilization feedback.
\begin{itemize}
	\item 
	\emph{RB scale-up action denoted $a'_1$:} We define $\Omega_d^{c,k}=\frac{\tilde{B}_d^{c,k}}{B_d^{c,k}}$ when  $\Psi_d^{c,k} = B_d^{c,k}$ and $\varphi^{c,k} < 1$. In this scenario, there are numerous incoming packets for slice $c$ associated with service $k$. However, slice $c$ is unable to meet the delay budget requirement. Closed-loop $1$ performs slice resource scale-up for slice $c$.
	\item  
	\emph{RB scale-down action denoted $a'_2$:} We use $\Omega_d^{c,k}=\frac{B_d^{c,k}}{\tilde{B}_d^{c,k}}$ if  $\Psi_d^{c,k} > B_d^{c,k}$ and  $\tilde{\varphi}^{c,k}_d < 1$. In this scenario, RBs assigned to O-DU $d$ are underutilized because $E[\lambda_{v}^{k,p}]$ is small. The close-loop $1$ performs RB scale-down for slice $c$. 
	\item 
\emph{RB allocation stopping action denoted $a'_3$:} When   $\Psi_d^{c,k} = \tilde{B}_d^{c,k}$,  we set $\Omega_d^{c,k}=0$. In this scenario, there is no demand for service $k$ associated with slice $c$ because $E[\lambda_{v}^{k,p}]= 0$. Close-loop $1$ can stop RB allocation to that slice and disable slice $c$.
	\item 
	\emph{Keep initial RB allocation action denoted $a'_4$:} When actions $a'_1$, $a'_2$, and $a'_3$ are not taken,  we set $\Omega_d^{c,k}=1$. The closed-loop $1$ assumes that the initial RB allocation is well performed, and thus, there is no requirement to update the RB allocation.
\end{itemize}
In our proposal, we use $\Omega_d^{c,k}$ to quantify the actions $a'_1$, $a'_2$, $a'_3$, and $a'_4$ of closed-loop $1$, where $\mathcal{A}'(\vect{y})= (a'_1, a'_2, a'_3,a'_4)$ is the action space. 
 
 \subsubsection{Zero-Touch Resource Management for Closed-loop $2$} 
 
The closed-loop $1$ shares $\Omega_d^{c,k}$ and $\mathcal{A}'(\vect{y})$ with the closed-loop $2$. Then,  the closed-loop $2$ can adjust the initial RB allocation for O-DUs. In other words, RB adjustment actions of closed-loop $1$ impact RB adjustment actions of closed-loop $2$ and vice-versa. Also, closed-loop $1$ should send feedback to closed-loop $2$ showing the difference $\nu^c_d=\beta^{c,k}_d- \lfloor\Omega_d^{c,k} \sum_{v\in \mathcal{V}^k}y_{v,m}^{\beta}\beta_v \rfloor
$ between RB demands from CPEs $\lfloor\Omega_d^{c,k} \sum_{v\in \mathcal{V}^k}y_{v,m}^{\beta}\beta_v \rfloor$ and the  allocated RBs $\beta^{c,k}_d$ to O-DU. The RB management actions for close-loop $2$ can be expressed as follows:
\begin{itemize}
	\item 
	\emph{RB scale-up action denoted $a_1$:} When $\nu^c_d<0$, closed-loop $2$ performs RB scale-up for slice $c$ using $\Omega_d^{c,k}$ from closed-loop $1$.
	\item  
	\emph{RB scale-down  action denoted $a_2$:} If RBs $\nu^c_d>0$, closed-loop $2$ performs RB scale-down for slice $c$ with $\Omega_d^{c,k}$ from closed-loop $1$. 
	\item 
	\emph{RB allocation stopping action denoted $a_3$:} When $ \nu^c_d=\beta^{c,k}_d$, closed-loop $2$ terminates RB allocation for slice $c$ with $\Omega_d^{c,k}=0$ because RB assigned to slice $c$ are not used, i.e., $\lfloor\Omega_d^{c,k} \sum_{v\in \mathcal{V}^k}y_{v,m}^{\beta}\beta_v \rfloor=0$.
	\item 
	\emph{Keep initial RB allocation action (denoted $a_4$):} When actions $a_1$, $a_2$, and $a_3$ are not taken,   closed-loop $2$ uses  $\Omega_d^{c,k}=1$ and keeps initial RBs allocation unmodified because it was well performed.
\end{itemize}
Here, $\nu^c_d$ and  $\Omega_d^{c,k}$ link actions of closed-loop $2$ with actions of closed-loop $1$. Furthermore, we use $\mathcal{A}(\vect{x}, \vect{y})= (a_1, a_2, a_3,a_4)$ as action space of closed-loop $2$.

\subsection{Energy Model for Serving Edge Cloud}
\label{subsec:EnergyModel}

At edge cloud, we consider $P$ as the number of servers. The total power of each server $p \in \mathcal{P}$ is given by $p^{pow}=  p^{base} + \sum_{w=1}^{W}  p_{v,m}^w$, 
where $W$ is the number of virtual instances such O-DUs and O-CU-UP at  server $p$, and 
$p_{v,m}^w$ is the power of active virtual instances. We use $p^{pow}$ as the total power of a physical server, while $p^{base}$ is the baseline power that is empirically determined. Based on arrival rate $\lambda_{v}^{k,p}$ and service rate  $\mu_{v}^{k,p}$, the total power consumption $\pi(P)$ becomes:
\begin{equation}
	\label{eq:energy_requirement_satisfaction}
	\pi(P) =\sum_{p=1}^{P}( p^{pow} \sum_{v=1}^{V_p}  \frac{y_{v,m}^{\beta}\lambda_{v}^{k,p}}{W\mu_{v}^{k,p}}) \, ,
\end{equation}
where $V_p \leq V $ is the number of CPEs served by server $p$. Furthermore, let us consider $T$ as a large timescale, such as a day, subdivided into small timescales $t$. During the time $t$, by considering energy and power measurements,  total  energy consumption $L_{cons}(t)$ becomes:
\begin{equation}
	\label{eq:energy_total}
	L_{cons}(t) = \pi(P) t.
\end{equation}
The equations ($\ref{eq:energy_requirement_satisfaction}$) and (\ref{eq:energy_total}) show that switching off the unused server and disabling unused virtual instances such as O-DU can help in minimizing the energy consumption of the edge cloud. However, managing energy for  Heating, Ventilation, and Air Conditioning (HVAC)\cite{petri2021edge} is out of the scope of this paper.

Let us denote $L(t)$  as total available energy at time $t$,
where the energy $L(t)$ comes from power grid $h^{+}_t$ and renewable energy  $g_t$. Also, we have energy storage. The total energy $L(t)$ to serve the edge cloud can be expressed as follows:
\begin{equation}
	\label{eq:enegy_satisfaction}
	L(t) = t	(( (1- z_{\Phi,t})\,((\,g_t +  h^{+}_t) -\Phi^{+}_t)) \,+ 	z_{\Phi,t}	\Phi^{-}_t) \,,
\end{equation}
where $\Phi^{-}_t$ represents energy discharged from energy storage to serve the edge cloud and $\Phi^{+}_t$ denotes the energy storage charging. Here, to ensure  energy storage is not empty, we use $z_{\Phi,t}$ as decision variables such that:
\begin{equation}
	\setlength{\jot}{10pt}
	z_{\Phi,t} =
	\begin{cases}
		1 \text{ if renewable energy is discharged from storage},\\
		0,\;\text{otherwise (i.e., energy from power grid is used).}
	\end{cases}
\end{equation}
Furthermore, to ensure that the edge cloud has enough energy to serve CPEs and O-RAN elements, we impose the following constraint:
\begin{equation}
	\label{eq:enegy_constraint2}
	(z_{\Phi,t} + ( \, 1-z_{\Phi,t} ))(x^{c,k}_{d} + y_{v,m}^{\beta}) > 1.\\
\end{equation}
Also, to guarantee that energy consumption should be less than the available total energy, we formulate the following constraint: 
\begin{equation}
	\label{eq:enegy_requirement}
	L_{cons}(t) \leq L(t).
\end{equation}

We assume that the maximum amount of energy that the edge cloud can obtain from the energy grid and renewable energy sources is limited at each time $t$. The maximum energy amounts are $h^{max}$ for the power grid and  $g^{max}$ for renewable energy. Therefore, the power grid and renewable energy should satisfy the following constraints:
\begin{equation}
	\label{eq:enegy_1}
	0	\leq  (1-z_{\Phi,t}) h^{+}_t \leq h^{max}.
\end{equation}
\begin{equation}
	\label{eq:enegy_2}
	0	\leq z_{\Phi,t} g_t  \leq g^{max}.
\end{equation}
Also, the energy storage has limited capacity with maximum amounts of energy to recharge $\Phi^{+}_{max}$ and discharge $\Phi^{-}_{max}$. Therefore, energy charging and discharging should meet the following constraints:
\begin{equation}
	\label{eq:enegy_5}
	0 \leq  z_{\Phi,t} \Phi^{+}_t \leq \Phi^{+}_{max},
\end{equation}
\begin{equation}
	\label{eq:enegy_6}
	0 \leq z_{\Phi,t}\Phi^{-}_t \leq \Phi^{-}_{max}.
\end{equation}

We assume renewable energy is harvested for free. When renewable energy generation is high and $    L(t)> L_{cons}(t)$,  the surplus energy denoted	$h^{-}_{t+1}$ can be sold to the energy market at next time $t+1$. The surplus energy 	$h^{-}_{t+1}$ can be defined as follows:
\begin{equation}
	\setlength{\jot}{10pt}
	h^{-}_{t+1} =
	\begin{cases}
	L(t) -	L_{cons}(t),\; \text{if  $L(t) >L_{cons}(t)$},\\
		0,\;\text{otherwise.}
	\end{cases}
\end{equation}
We consider $h^{-}_{t+1}$ is in energy storage. Therefore, energy storage must satisfy  the following  capacity constraint:
\begin{equation}
	\label{eq:enegy_7}
	h^{-}_{t+1} + \Phi^{-}_t\leq \Phi^{+}_t \leq \Phi^{+}_{max}.
\end{equation}
Furthermore, we define energy cost  $H(t)$ for edge cloud provider as follows: 
\begin{equation}
	\label{eq:enegy_satisfaction2}
	H(t) =   h^{+}_t \tilde{\zeta}_E  -	h^{-}_{t+1} {\zeta}_E \,.
\end{equation}
Let $\tilde{\zeta}_E$ represent the cost of grid energy from the energy provider during the timescale $t$, and ${\zeta}_E$ denote the selling price of surplus energy $h^{-}_{t+1}$ for the timescale $t+1$, as determined by the edge cloud provider. We assume ${\zeta}_E \leq \tilde{\zeta}_E$ because if ${\zeta}_E \geq \tilde{\zeta}_E$, there would be no financial incentive to purchase the surplus energy from the edge cloud provider. Here, the energy costs are expressed in monetary terms.

\section{Problem Formulation for Physical Twin} 
\label{sec:ProblemFormulation}
A time $t$ for the energy model, such as an hour, is large for the communication model. Therefore, we define the communication model's ultra-small timescale $s$ in milliseconds (ms). Then, we formulate an ultra-small timescale optimization problem that maximizes slice requirement satisfaction and RB utilization at O-DUs as follows:
\begin{subequations}
	\label{eq:problem_formulation1}
	\begin{align}
		&\underset{(\vect{x},\vect{y})}{\text{max}}\ \  \sum_{k=1}^{K_d}(x^{c,k}_{d} (s)\tilde{\varphi}^{c,k}_d(s) + \varphi^{c,k}(s))
		\tag{\ref{eq:problem_formulation1}}\\
		& \text{subject to: }\nonumber\\
		&\sum_{k=1}^{K_d}x^{c,k}_{d}(s) \beta^{c,k}_d(s)\leq \beta_d(s)\label{first:a1},\\
		& \lfloor\Omega_d^{c,k}x^{c,k}_{d}(s)\sum_{v \in \mathcal{V}^k}^{}y_{v,m}^{\beta} (s)\beta_v (s) \rfloor\leq \beta^{c,k}_d(s),\label{first:b1}\\
		&	\sum_{v \in \mathcal{V}^k}^{}\lambda^v_{k,c}y_{v,m}^{\beta} (s) R^{c,k}_v(s) \, \leq\varpi_{m,d}(s), \label{first:c1}\\
		& \sum_{u\in \mathcal{V}^k}x^{c,k}_{d}(s)\leq 1, \; \label{first:d1}\\
		&\sum_{v\in \mathcal{V}^k}y_{v,m}^{\beta}(s)\leq 1 \label{first:e1}.
	\end{align}
\end{subequations}

In the formulated ultra-small scale optimization problem (\ref{eq:problem_formulation1}), the constraint in
(\ref{first:a1})  guarantees that the RB allocation to slice $c$ associated with service $k$ does not exceed the available RB of O-DU. Constraint (\ref{first:b1})  ensures that the RB allocation to CPEs does not exceed the available RB of each O-DU $d$. 
The constraint in (\ref{first:c1}) concerns the fronthaul network, ensuring that CPEs do not transmit more traffic than the fronthaul capacity allows. The constraint in (\ref{first:d1}) ensures that each slice $c$ associated with service $k$ is at one O-DU (no duplication). The constraint in (\ref{first:e1}) ensures that RB $\beta_v(s)$ can only be allocated to a single CPE.

Solving (\ref{eq:problem_formulation1}) is challenging because it is a non-convex problem. Also, closed-loop $1$ and closed-loop $2$ operate in different ultra-small timescales. Therefore,  we split (\ref{eq:problem_formulation1}) into two sub-problems for closed-loop $1$ (ultra-small timescale less than $10$ ms) and for closed-loop $2$ (ultra-small timescales within the range from $10$ to $1000$ ms).

The ultra-small timescale optimization problem for closed-loop 2 that maximizes expected RB  utilization at O-DU can be expressed as follows:
\begin{subequations}
	\label{eq:problem_formulation2}
	\begin{align}
		&\underset{(\vect{x})}{\text{max}}  \frac{1}{s_2}\sum_{s={10}}^{s_2}\mathop{{}\mathbb{E}} \{ \sum_{k=1}^{K_d}x^{c,k}_{d}(s)\tilde{\varphi}^{c,k}_d (s) \}
		\tag{\ref{eq:problem_formulation2}}\\
		& \text{subject to: }\nonumber\\
		& \sum_{k=1}^{K_d}x^{c,k}_{d}(s) \beta^{c,k}_d(s)\leq \beta_d(s),\\
		& \sum_{u\in \mathcal{V}^k}x^{c,k}_{d}(s)\leq 1, \; \label{first:c2}
	\end{align}
\end{subequations}
where $s2<1000$ $ms$ is the time limit for closed-loop $2$ to distribute RBs to O-DUs. We consider Near-RT RIC handles (\ref{eq:problem_formulation2}), and it shares information with O-DU to solve the following ultra-small timescale optimization problem for closed-loop $1$:
\begin{subequations}
	\label{eq:problem_formulation3}
	\begin{align}
		&\underset{(\vect{x}, \vect{y})}{\text{max}} \frac{1}{s_1}\sum_{s={1}}^{s_1}\mathop{{}\mathbb{E}} \{\sum_{k=1}^{K_d} \varphi^{c,k}(s)\}
		\tag{\ref{eq:problem_formulation3}}\\
		& \text{subject to: }\nonumber\\
		& \lfloor\Omega_d^{c,k}x^{c,k}_{d}(s)\sum_{v \in \mathcal{V}^k}^{}y_{v,m}^{\beta} (s)\beta_v (s) \rfloor\leq \beta^{c,k}_d(s),\label{first:a11}\\
		&	\sum_{v \in \mathcal{V}^k}^{}\lambda^v_{k,c}y_{v,m}^{\beta} (s) R^{c,k}_v(s) \, \leq\varpi_{m,d}(s),\label{first:b11}\\
		&\sum_{v\in \mathcal{V}^k}y_{v,m}^{\beta}(s) \leq 1 \label{first:d11},
	\end{align}
\end{subequations}
where $s1<$ $10ms$ denotes the time limit for closed-loop $1$ to allocate RBs to CPEs. The problem (\ref{eq:problem_formulation3}) maximizes expected slice requirement satisfaction in terms of delay and depends on (\ref{eq:problem_formulation2}). In other words, those two closed-loops need to share information, where variable $x^{c,k}_{d}(s)$ links them.

By applying optimization approaches for solving (\ref{eq:problem_formulation2}) and (\ref{eq:problem_formulation3}), we may obtain stationary solutions that are not suitable for RB auto-scaling, as the resource auto-scaling processes are dynamic and not stationary tasks. Therefore,  to have zero-touch resource adjustment, we change  (\ref{eq:problem_formulation2}) and (\ref{eq:problem_formulation3}) to reward functions.  We define a reward function $r_{s,d}(\vect{x})$ for closed-loop $2$ to assess the utilization of RB $\beta_d$ at O-DU $d$ at an ultra-small timescale $s$.
\begin{multline}
	\label{eq:problem_formulation22}
	r_{s,d}(\vect{x})=  \frac{1}{s_2}\sum_{s={10}}^{s_2}\mathop{{}\mathbb{E}} \{ \sum_{k=1}^{K_d}x^{c,k}_{d}\tilde{\varphi}^{c,k}_d (s) \} +  	\Delta_d(1-\sum_{u\in \mathcal{V}^k}x^{c,k}_{d}(s))+ \\ \Delta_{\beta}( (\beta_d(s)-\sum_{k=1}^{K_d}x^{c,k}_{d}(s) \beta^{c,k}_d(s))+ \nu^c_d).
\end{multline}
where the penalty parameter $\Delta_d$ ensures that each slice is managed by only one O-DU. We denote the penalty $\Delta_{\beta}$ to ensure that RB updates do not violate the RB constraint. Initially, the closed loop can use $\nu^c_d=0$. Then, it can update $\nu^c_d$ after receiving the closed-loop $1$ feedback.

We define a reward function $r_{s,c}(\vect{y})$ for closed-loop $1$ to reflect the satisfaction of slice requirement and workload changes at an ultra-small timescale $s$:
\begin{multline}
	\label{eq:problem_formulation11}
	r_{s,c}(\vect{y})=\frac{1}{s_1}\sum_{s={1}}^{s_1}\mathop{{}\mathbb{E}} \{\sum_{k=1}^{K_d} \varphi^{c,k}(s)\} +	\Delta_m(\varpi_{m,d}(s)-	\\ \sum_{v \in \mathcal{V}^k}^{}\lambda^v_{k,c}y_{v,m}^{\beta} (s) R^{c,k}_v(s)) + 	\Delta_v(1- \sum_{u\in \mathcal{V}^k}y_{v,m}^{\beta}) +  \Delta_z\nu^c_d.
\end{multline}
We denote $\Delta_m$ as the penalty for violating the fronthaul resource constraint. $\Delta_v$ serves as the penalty parameter for violating the RB allocation constraint, while $\Delta_z$ is the penalty parameter to ensure that intra-slice scaling does not allocate more RBs than the total available RBs.

Closed-loop $1$ aims to maximize the reward function $r_{s,c}(\vect{y})$ by meeting the intra-slice delay budget constraint and managing workload changes at an ultra-small timescale $s$. Conversely, closed-loop $2$ seeks to maximize its reward $r_{s,d}(\vect{x})$  by avoiding violations of the RB capacity constraints at each O-DU. However, the utilization of RBs at O-DU depends on the intra-slice RB allocation. Thus, $\nu^c_d$ links the actions of closed-loop $1$ with those of closed-loop $2$. Closed-loop 1 must transmit $\nu^c_d$ to closed-loop $2$ as feedback, indicating the difference between RB demands and the RBs allocated to O-DU. Therefore, we define a main reward function $r_s(\vect{x},\vect{y})$ that connects these two proposed closed-loops at an ultra-small timescale $s$, where $r_s(\vect{x},\vect{y})$ is given by:
\begin{equation}
	\label{eq:problem_formulation33}
	r_s(\vect{x},\vect{y})=\phi_{dis} 	r_{s,d}(\vect{x}) + (1-\phi_{dis}) 	r_{s,c}(\vect{y}).
\end{equation}
As closed-loop $2$ maximizes the reward in (\ref{eq:problem_formulation33}), which combines (\ref{eq:problem_formulation22}) and (\ref{eq:problem_formulation11}), and (\ref{eq:problem_formulation11}) is also maximized with closed-loop $1$, we introduce $\phi_{dis}$ as a discount parameter. This parameter allows closed-loop $2$ to balance the trade-off between maximizing rewards (\ref{eq:problem_formulation22}) and (\ref{eq:problem_formulation11}).

\subsection{A Joint Problem for Communication and Energy Models}
An edge cloud that hosts O-RAN elements for allocating radio resources to CPEs requires energy. Consequently, it is necessary to integrate the communication model with the energy model. We propose the following small-scale optimization problem, defined in terms of $t$, which aims to minimize energy costs while maximizing communication utility:
\begin{subequations}
	\label{eq:problem_formulation4}
	\begin{align}
		&\underset{(\vect{x},\vect{y}, \vect{z})}{\text{min}}\ \   H(t) - \frac{1}{t}{\zeta}_c\sum_{s= 1000}^{t}\mathop{{}\mathbb{E}} \{\sum_{d=1}^{D}\beta^{c,k}_d(s)\}
		\tag{\ref{eq:problem_formulation4}}\\
		& \text{subject to: (\ref{eq:enegy_constraint2}) - (\ref{eq:enegy_6}), and (\ref{eq:enegy_7}).}
	\end{align}
\end{subequations}
In (\ref{eq:problem_formulation4}),  ${\zeta}_c$ is fixed selling price of $\beta^{c,k}_d$ for timescale $t$. In other words, $\frac{1}{t}{\zeta}_c\sum_{s= 1000}^{t}\mathop{{}\mathbb{E}} \{\sum_{d=1}^{D}\beta^{c,k}_d(s)\}$ is communication utility. 

The Near-RT RIC can share information from closed-loop $2$, such as $\beta^{c,k}_d(s)$ and ${\zeta}_c$, for the period $t$ with the Non-RT RIC. Additionally, the energy controller can communicate the energy cost $H(t)$ to the Non-RT RIC via the Near-RT RIC. Subsequently, the Non-RT RIC manages Equation (\ref{eq:problem_formulation4}) outside of closed-loops $1$ and $2$, as $t$ is measured in terms of an hour. Furthermore, the Non-RT RIC can transmit historical solution data to the MEC server for constructing or updating the Digital Twin (DT), as described in the subsequent section. With the aid of machine learning, the DT can project up to $\tau$ hours ahead and provide feedback on Resource Block (RB) allocation and energy information to the Near-RT RIC for a prolonged time window. This allows closed-loops $1$ and $2$ to manage their rewards during the next ultra-small timescales.

Closed-loop $2$  takes  actions $\mathcal{A}(\vect{x}, \vect{y})$ of allocating initial RB, keeping initial RB allocation ($\nu^c_d=0$), RB scaling-up ($\nu^c_d>0$), RB scaling-down ($\nu^c_d<0$) , and terminating RB allocation for O-DU ($ \nu^c_d=\beta^{c,k}_d$, i.e., $\lfloor\sum_{u\in \mathcal{V}^k}y_{v,m}^{\beta}\beta_v\Omega_d^{c,k}\rfloor=0$).  The states at Near-RT RIC, denoted as $\mathcal{S}= {(\vect{\beta}, \vect{D}, \vect{C})}$, encompass the states of RBs ($\vect{\beta}$), O-DUs ($\vect{D}$), and slices ($\vect{C}$) managed by O-DUs. On the other hand, closed-loop 1 takes actions $\mathcal{A}'(\vect{y})$ involving the assignment of initial RBs, maintaining the initial RBs allocation ($\Omega_d^{c,k}=1$), scaling-up RBs ($\Psi_d^{c,k} = \beta^d_{c,k}$), scaling-down RBs ($\Psi_d^{c,k} > \beta^d_{c,k}$), and terminating RBs allocation ($\Psi_d^{c,k} = \tilde{\beta}^d_{c,k}$) for CPEs. The states $\mathcal{S}'= {( \vect{V}, \vect{\Omega},\vect{\Psi}) }$ consist of the states of $\vect{V}$ CPEs managed by slices, intra-slice orchestration ($\vect{\Omega}$), and queue status ($\vect{\Psi}$). Closed-loop 1 observes demands from CPEs and assigns RBs to them based on queue status and intra-slice requirement satisfaction. It can then maintain or update the RBs allocation for CPEs. Subsequently, it provides feedback to closed-loop 2, allowing closed-loop 2 to maximize (\ref{eq:problem_formulation22}) and (\ref{eq:problem_formulation33}).

\section{Proposed Solution and Digital Twin Assistance} 
\label{sec:ProposedSolution}

\begin{figure}[t]
	\centering
	\includegraphics[width=1.0\columnwidth]{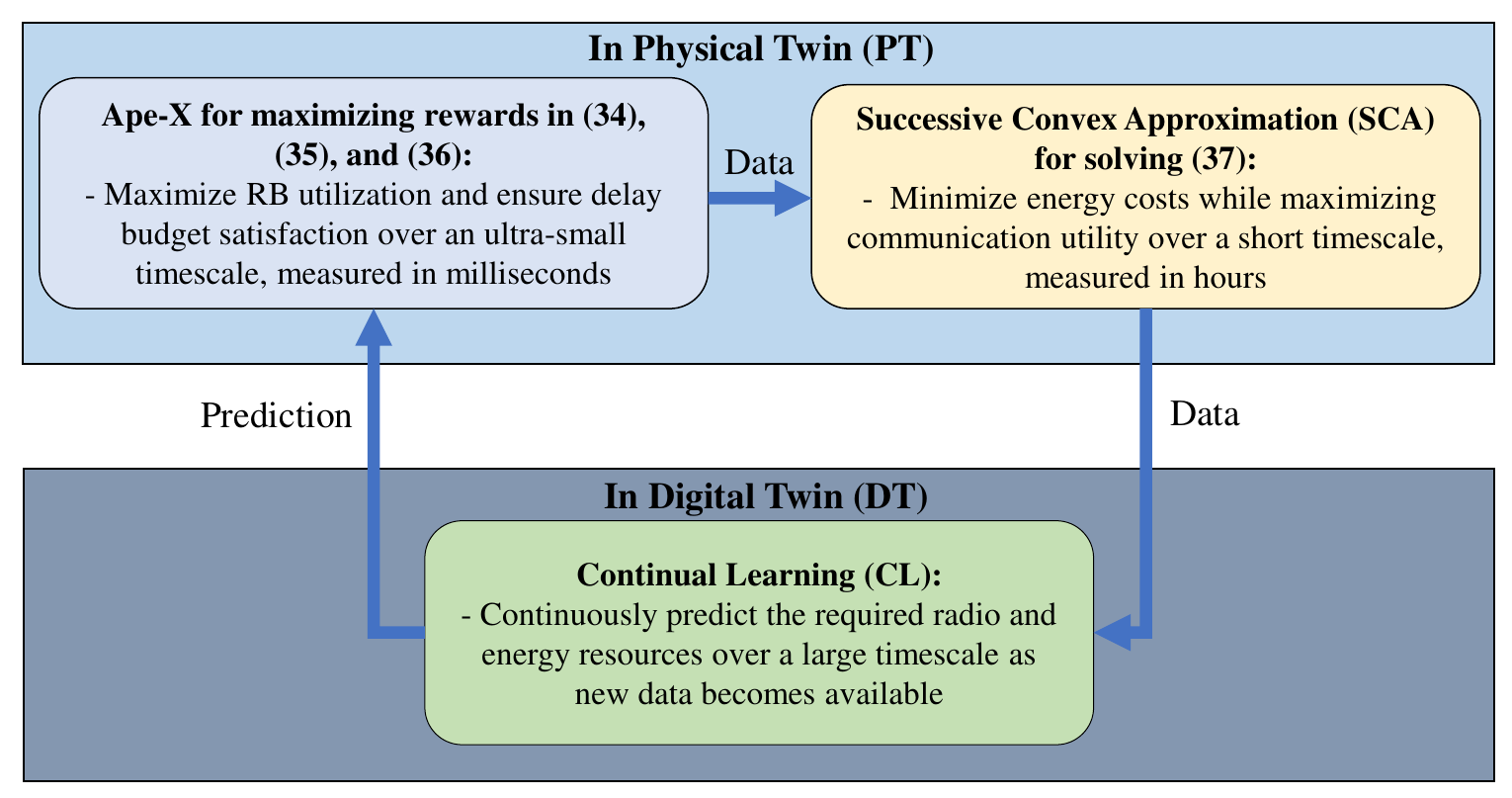}
	\caption{Summary of the proposed solution.}
	\label{fig:SolutionApproachTMC}
\end{figure}

As summarized in Fig. \ref{fig:SolutionApproachTMC}, in this section, we discuss the Ape-X algorithm for maximizing the formulated rewards of two closed-loops and Successive Convex Approximation (SCA) for solving joint energy and communication optimization in the physical twin (PT). We then explore Continual Learning (CL) in the digital twin (DT), which leverages data from the PT to predict the required energy and communication resources.
\begin{figure*}[t]
	\centering
	\includegraphics[width=2.0\columnwidth]{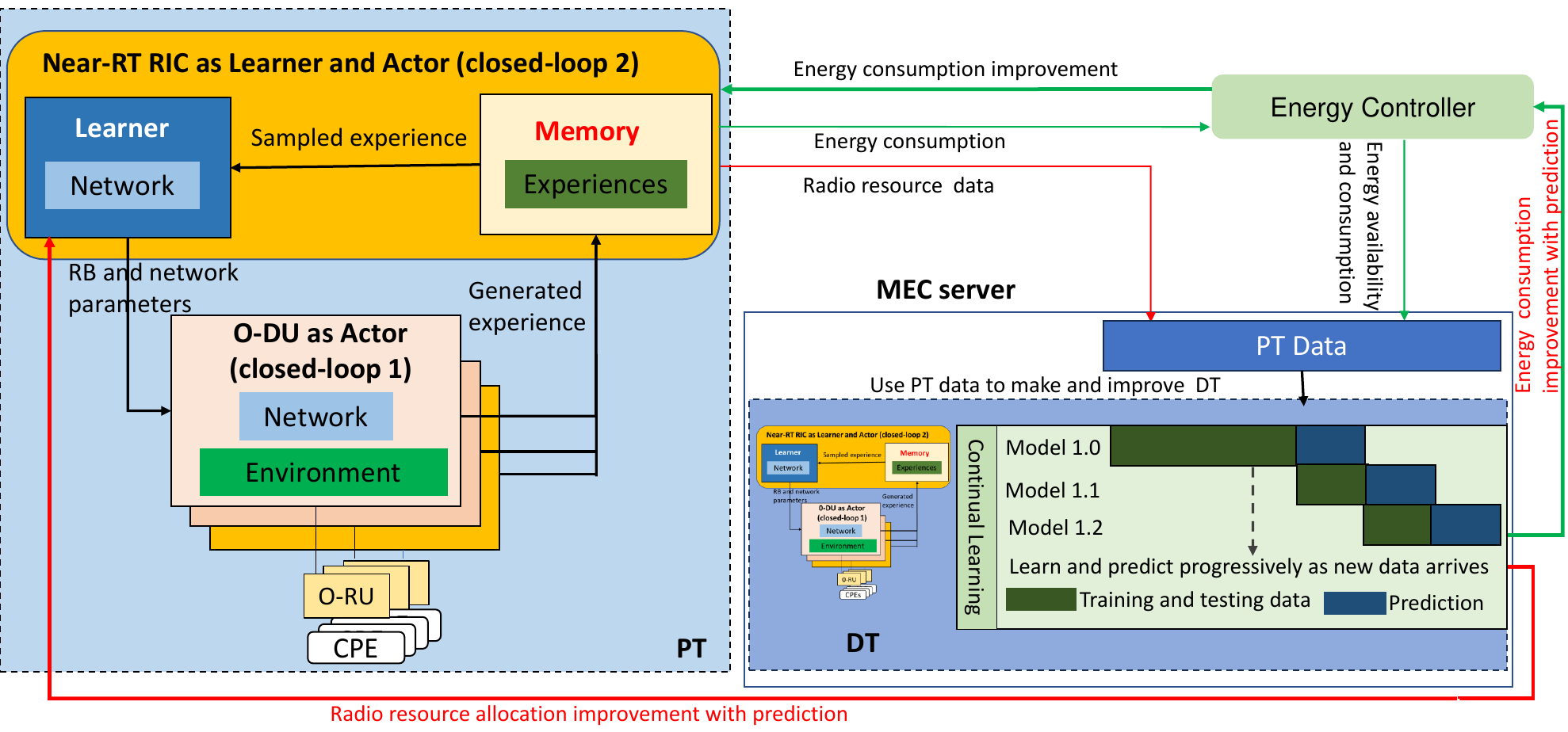}
	\caption{The Ape-X in PT  and CL in DT.}
	\label{fig:Apex}
\end{figure*}

In PT, we use Reinforcement Learning (RL) \cite{kiran2021deep} to maximize the formulated rewards in (\ref{eq:problem_formulation22}), (\ref{eq:problem_formulation11}), and (\ref{eq:problem_formulation33}). We choose Ape-X \cite{horgan2018distributed} because this distributed RL algorithm can be executed in the Near-RT RIC and O-DUs, as illustrated in Fig. \ref{fig:Apex}.
Ape-X divides deep reinforcement learning into two key components. The first component, referred to as the actor, interacts with the environment, implements and evaluates the shared Deep Neural Network (DNN) model, and stores observation data in replay memory. The second component, known as the learner, uses batches of data from the replay memory to update the DNN model and adjust learning parameters.

 In our approach, we consider O-DUs as an actor for closed-loop $1$ that interacts with CPEs for maximizing (\ref{eq:problem_formulation11}) and Near-RT RIC as an actor for closed-loop two that interacts with O-DUs for maximizing (\ref{eq:problem_formulation22}) and (	\ref{eq:problem_formulation33}). Also,  Near-RT RIC  acts as a learner for updating learning parameters. In Ape-X,  the loss function $\Phi_j(\theta)$ to be minimized can be defined as follows:
\begin{equation}
	\begin{aligned}
		\label{eq:loss2}
		\Phi_j(\theta)=\frac{1}{2}(\tilde{G}_j- Q(s_j, a_j,\theta))^2,
	\end{aligned}
\end{equation}
where $Q(s_j, a_j,\theta)$ is action-value estimate and $\theta$ is parameters of the DNN. We use
$\tilde{G}_j$ to denote the return function, where $\tilde{G}_j$ is given by:
\begin{equation}
	\begin{aligned}
		\label{eq:problem_discounted rewards}
		\tilde{G}_t= r_{j+1} + \gamma r_{j+2}+ \dots + \gamma^{n-1} r_{j+n} + \\ \gamma^{n} Q(s_{j+n}, \underset{\vect{a} }{\text{argmax}} Q(s_{j+n}, a, \theta) , \theta^{-}).
	\end{aligned}
\end{equation}
We use $n$ to denote the number of steps, and $j$ is a time index of sampling experience in replay memory. Here, we consider Near-RT RIC and replay memory to be on the same server at the edge cloud. The experience sampling considers state $s_j \in \mathcal{S}|\mathcal{S}'$, actions $a, a_j\in \mathcal{A}| \mathcal{A}'$, and parameters of the target network $\theta^{-}$.

After obtaining $\vect{x}$ and $\vect{y}$ using Ape-X in the ultra-small timescale $s$, we use $\vect{x}$ and $\vect{y}$  to determine the solution of the problem in (\ref{eq:problem_formulation4}) in the small scale $t$ for energy. For solving (\ref{eq:problem_formulation4}), we use SCA that applies Majorization-Minimization (MM) technique described in \cite{scutari2018parallel} and summarized as follows:
\begin{itemize}
\item 
In the Majorization step, we employ quadratic penalization to establish a convex proximal function, which serves as an upper bound of (\ref{eq:problem_formulation4}).
\item 
In the Minimization step, we aim to minimize the proximal upper-bound function instead of directly minimizing (\ref{eq:problem_formulation4}), ensuring that the upper-bound function takes steps proportional to the negative gradient.
\end{itemize}

In Majorization, let us use $\iota$ as iteration and $\mathcal{J}$ as a set of indexes in (\ref{eq:problem_formulation4}). We define  $\mathcal{F}(\vect{x},\vect{y}, \vect{z}) = H(t)-{\zeta}_c \frac{1}{t}\sum_{s= 1000}^{t}\mathop{{}\mathbb{E}} \{\sum_{d=1}^{D}\beta^{c,k}_d(s)\}$ and formulate the following proximal upper-bound function $\mathcal{F}_j(\vect{z_j},\vect{x}, \vect{y})$ for $j \in \mathcal{J}$ by adding a quadratic penalization to $\mathcal{F}(\vect{x},\vect{y}, \vect{z})$:
\begin{equation}
	\begin{aligned}
		\mathcal{F}_j(\vect{z_j},\vect{z}^{(\iota)}, \vect{x}^{(\iota)}, \vect{y}^{(\iota)})\defeq \mathcal{F}{(\vect{z}_j,\vect{\tilde{z}}, \vect{\tilde{x}}, \vect{\tilde{y}}}) + \frac{ \varrho_j}{2} \lVert(\vect{z}_j- \vect{\tilde{z}})\rVert^2. 
	\end{aligned}
	\label{eq:optimization_bsum1}
\end{equation}

\begin{figure}[t]
	\centering
	\includegraphics[width=1.0\columnwidth]{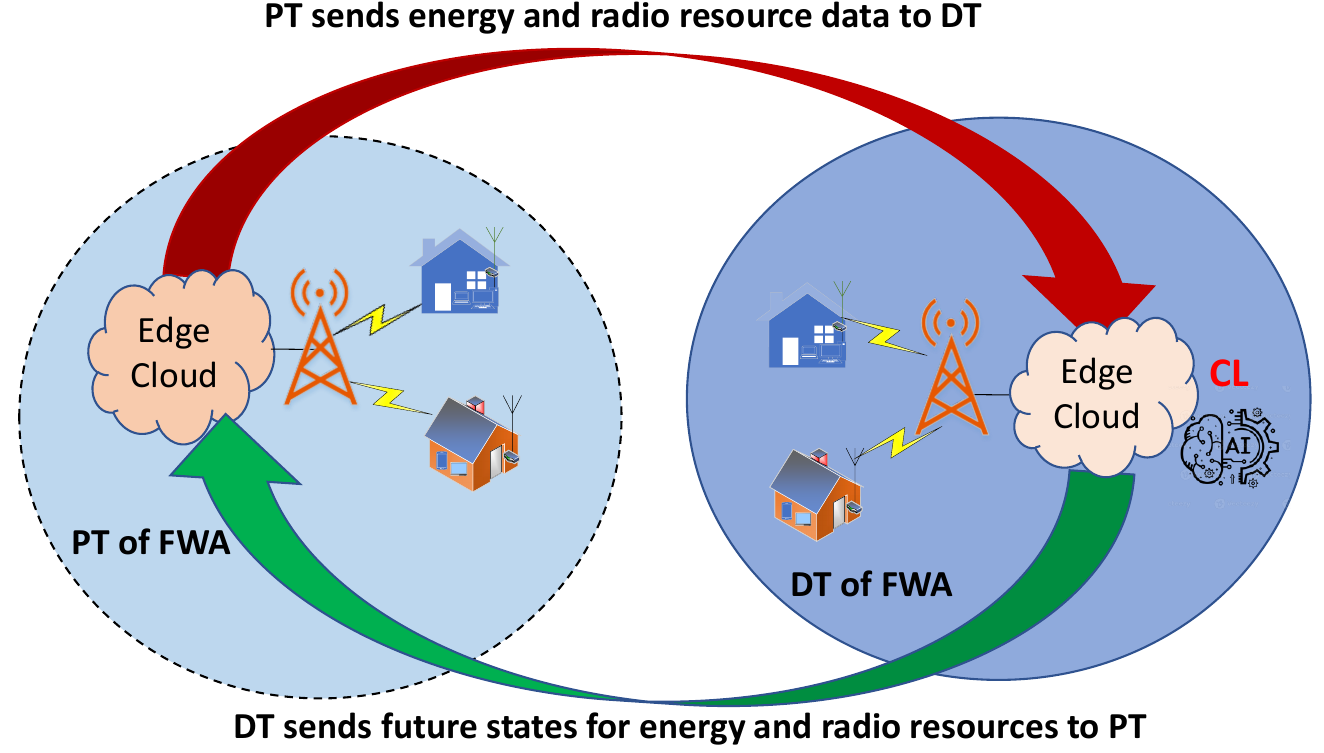}
	\caption{Illustration of solution-aware digital twins.}
	\label{fig:digitaltwins}
\end{figure}
The quadratic penalization $ \frac{ \varrho_j}{2} \lVert(\vect{z}_j- \vect{\tilde{z}})\rVert^2$ make proximal upper-bound function in (\ref{eq:optimization_bsum1}) convex and upper-bound of (\ref{eq:problem_formulation4}), where $\varrho_j>0$ denotes the positive penalty parameter. In other words, considering $\vect{z}_j$, $\vect{x}^{(\iota)}$, and  $\vect{y}^{(\iota)}$, we have minimizers vectors $\vect{\tilde{z}}$, $\vect{\tilde{x}}$, and $\vect{\tilde{y}}$ at each iteration $\iota$. The solutions from the previous step ($\iota-1$) are regarded as the initial solution. At each iteration $\iota+1$, the solution is updated by solving the following optimization problem:
\begin{equation}
	\begin{aligned}
		\vect{z}_j^{(\iota+1)}\in \underset{ \vect{z}_j \in \mathcal{Z}}{\text{min}}\; \mathcal{F}_j (\vect{z}_j,\vect{z}^{(\iota)}, \vect{x}^{(\iota+1)}, \vect{y}^{(\iota+1)}),
	\end{aligned}
	\label{eq:optimization22}
\end{equation} 
where  $\vect{\tilde{x}}$, $\vect{\tilde{y}}$, $\vect{x}^{(\iota+1)}$, and $\vect{y}^{(\iota+1)}$ come from Ape-X solution obtained by maximizing the reward functions (\ref{eq:problem_formulation11}), (\ref{eq:problem_formulation22}), and (\ref{eq:problem_formulation33}). Furthermore,  we define $\mathcal{Z}\triangleq\{\vect{z}: 	z_{\Phi,t}   \in [0,1]\}$ as the feasible sets of  $\vect{z}$. When $\vect{z}_j$ is relaxed, the problem in (\ref{eq:optimization22}) becomes convex and easy to solve using a solver such as an operator splitting solver for quadratic programs \cite{stellato2020osqp}. 

\subsection{Digital Twin Assists Physical Twin}
We utilize a DT to monitor and enhance PT by continuously feeding data from the PT into the DT, we predict the required radio and energy resources using CL to enhance closed-loop performance. Given that closed-loops 1 and 2 operate on ultra-small timescales, integrating CL, which requires a larger timescale, is challenging within these  closed-loops. Therefore, we employ CL within the DT.

In our approach, after solving (\ref{eq:optimization22}), we  store and use $\{H_{1:T}\} $ to represent data for energy cost, where $H_{1:T}=(H_{1}, H_{2}, \dots, H_{T})$.  Furthermore, we store and use $\{\beta^{c,k}_{d,1:T}\} $ to represent data for RB distribution at each O-DU $d$, where $\beta^{c,k}_{d,1:T}=(\beta^{c,k}_{d,1}, \beta^{c,k}_{d,2}, \dots, \beta^{c,k}_{d,T})$. In other words, the data $\{H_{1:T}\} $ and $\{\beta^{c,k}_{d,1:T}\}$ contain historical data for energy cost and RBs scheduling of the PT. As presented in Fig. \ref{fig:digitaltwins},  $\{H_{1:T}\} $ and $\{\beta^{c,k}_{d,1:T}\} $ are sent to the MEC server via the Non-RT RIC for creating DT of PT. In other words, the MEC server uses data and the network topology to make DT  and simulate PT, where information ties the PT and DT together. However, exchanging data between PT and DT consumes network energy:
\begin{equation}
L^{p}= (|H_{1:T}|+|\beta^{c,k}_{d,1:T}|) p_{pl}\Lambda f^2
\end{equation}
and bandwidth:
\begin{equation}
\varpi=\frac{|H_{1:T}|+|\beta^{c,k}_{d,1:T}|}{t_{second}},
\end{equation} 
where  $f$ represents the CPU frequency, and $\Lambda$ denotes the number of CPU cycles needed to send data sizes  $|H_{1:T}|$ and $|\beta^{c,k}_{d,1:T}|$. Furthermore, $p_{pl}$ is Planck's constant, and $t_{second}$  is time in second.

We define a time range $\{1,2, \dots, T$\} for the construction, updating, and DNN training phase for DT using network topology and data $\{H_{1:T}\} $ and $\{\beta^{c,k}_{d,1:T}\}$ of PT. Then, we  define time range $\{T+1, T+2, \dots, T+\tau$\} in the prediction phase. In other words, we predict  RBs $\{\tilde{\beta}^{c,k}_{d,{T+\tau}}\}$ and energy costs $\{\tilde{H}_{T+\tau}\}$ using DNN as future  resource states of PT, where $\tau$ is in terms of hours.  For a large timescale $\tau$, we minimize the Mean Square Error (MSE) between the predicted RBs and energy costs $\{\tilde{\beta}^{c,k}_{d,{T+\tau}},  \tilde{H}_{T+\tau}\}$ and ground truth RB allocation and energy costs $\{\beta^{c,k}_{d,1:T},  H_{1:T}\}$. As shown in  Fig. \ref{fig:Apex}, in this work, we choose  CL described in \cite{9833928} over other machine learning approaches because it allows the DNN model to learn progressively as new data $\{H_{1:T}\} $ and $\{\beta^{c,k}_{d,1:T}\}$ arrive. This continual learning approach is appropriate for dynamic scenarios, which do not require stationary data streams.

\begin{figure}[t]
	\centering
	\begin{minipage}{0.450\textwidth}
		\centering
		\includegraphics[width=1.0\columnwidth]{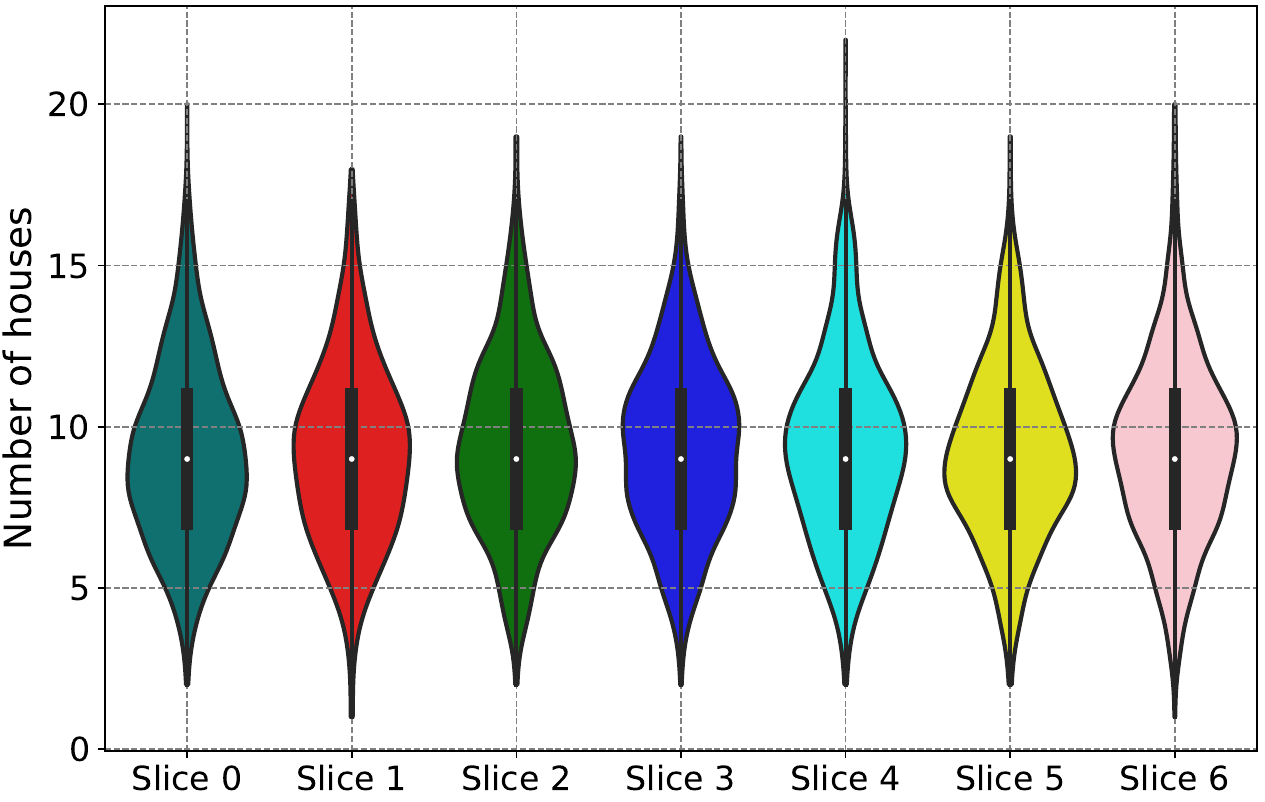}
		\caption{Number of CPEs per slice.}
		\label{fig:HouseSlice}
	\end{minipage}
	\begin{minipage}{0.450\textwidth}
		\centering
		\includegraphics[width=1.0\columnwidth]{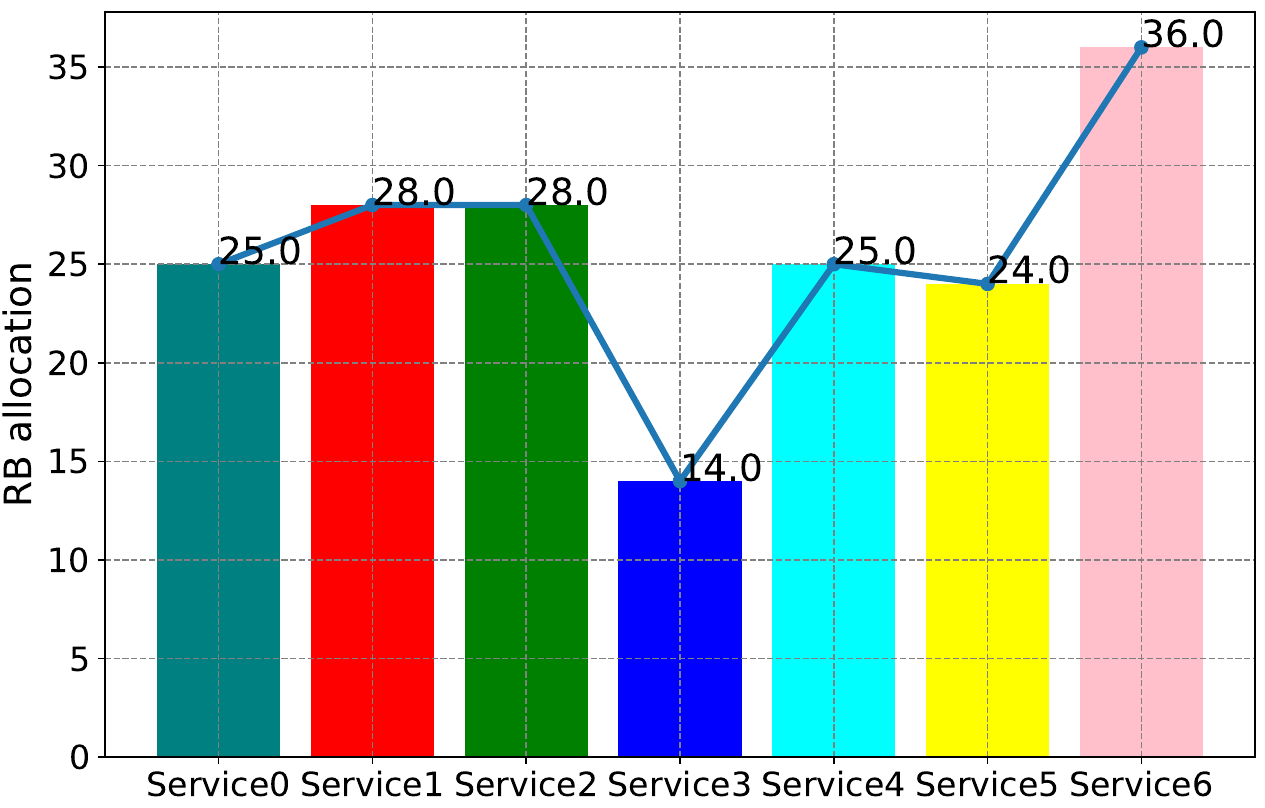}
		\caption{ RB for services.}
		\label{fig:Slice_vODU}
	\end{minipage}
		\begin{minipage}{0.45\textwidth}
		\centering
		\includegraphics[width=1.0\columnwidth]{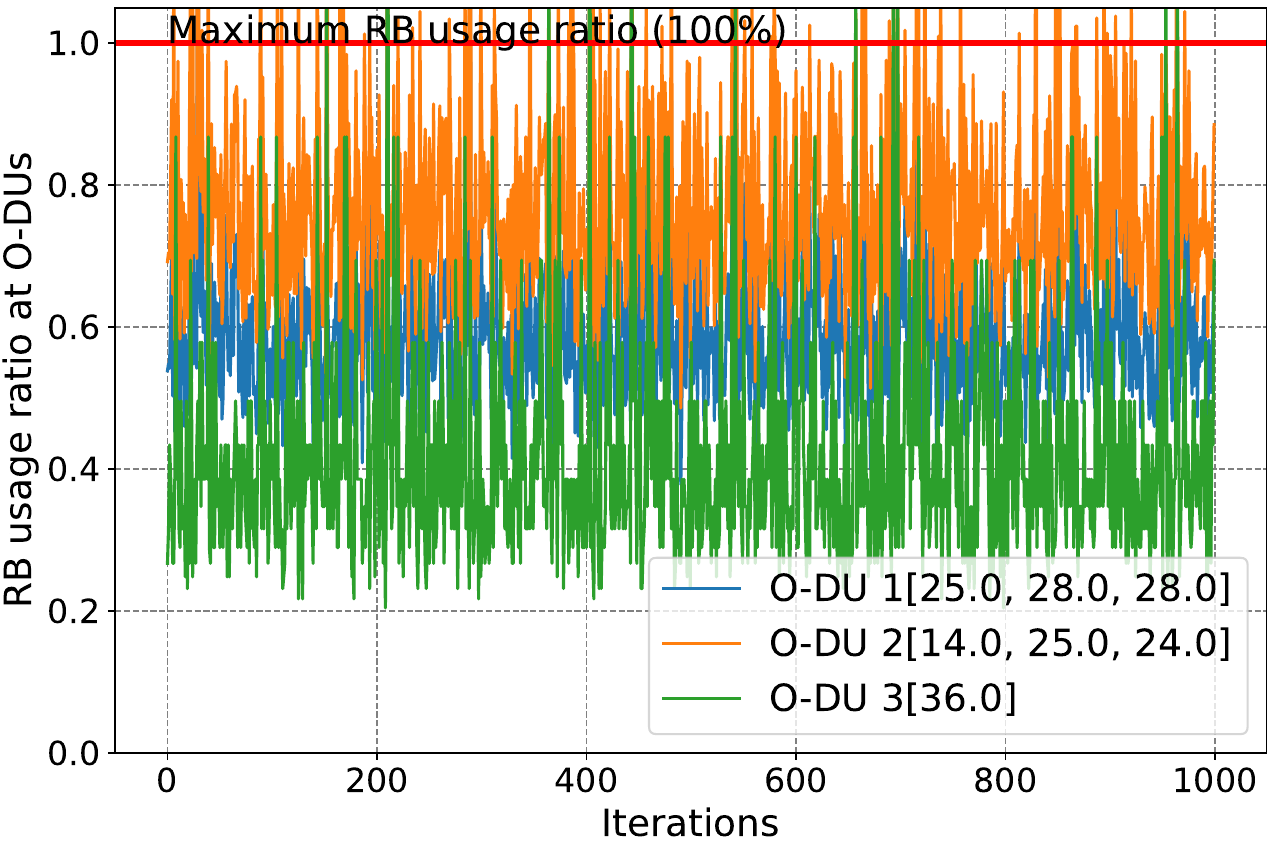}
		\caption{RB usage ratio (Eq. \ref{eq:RBusage}).}
		\label{fig:VDUUsage}
	\end{minipage}
\end{figure}
We consider a synchronization gap between PT and DT during the DT construction or updating, training, and prediction phases. Therefore, we define de-synchronization time $\Delta_\tau$  at the MEC server as the time needed for DT construction or updating, training, and prediction phases, where  $\Delta_\tau$ is given by:
\begin{equation}
	\label{eq:de-synchronization}
	\Delta_\tau=\frac{|H_{1:T}|+|\beta^{c,k}_{d,1:T}|}{f_{mec}}\Lambda_{mec}\iota_{mec},
\end{equation}
where $f_{mec}$ represents the CPU frequency of the MEC server, and $\iota_{mec}$ is the training iterations. $\Lambda_{mec}$ denotes the CPU cycles required in DT construction or update, training, and prediction phases.

DT construction or update, training, and prediction phases at the MEC server require the following
CPU energy:
\begin{equation}
	L^p_{mec}= (|H_{T+\tau}|+|\beta^{c,k}_{T+\tau}|) p_{pl}\Lambda_{mec} f_{mec}^2.
\end{equation} 
Furthermore, we formulate the following optimization problem to minimize the cost of  DT construction or update, training, and prediction phases:
\begin{subequations}
	\label{eq:problem_formulation6}
	\begin{align}
		&\underset{\vect{w}}{\text{min}}\ \  w^p_{mec} ( \tilde{\zeta}_E((L^{p} + L^p_{mec})\Delta_\tau) + \zeta \varpi)
		\tag{\ref{eq:problem_formulation6}}\\
		& \text{subject to: $  w^p_{mec}\leq 1$,}
	\end{align}
\end{subequations}
where ${\zeta}$ is the price of  bandwidth $\varpi$. We use $w^p_{mec}$ as binary decision variable, where $w^p_{mec}=1$ if MEC server receives data  $\{H_{1:T}\} $ and $\{\beta^{c,k}_{d,1:T}\}$ for DT. Otherwise, $w^p_{mec}=0$.  The formulated problem in (\ref{eq:problem_formulation6}) is an integer linear programming problem and can be solved using a solver such as Gurobi \cite{schuster2020exact}.
\begin{figure}[t]
	\centering
	\begin{minipage}{0.450\textwidth}
	\centering
	\includegraphics[width=1.0\columnwidth]{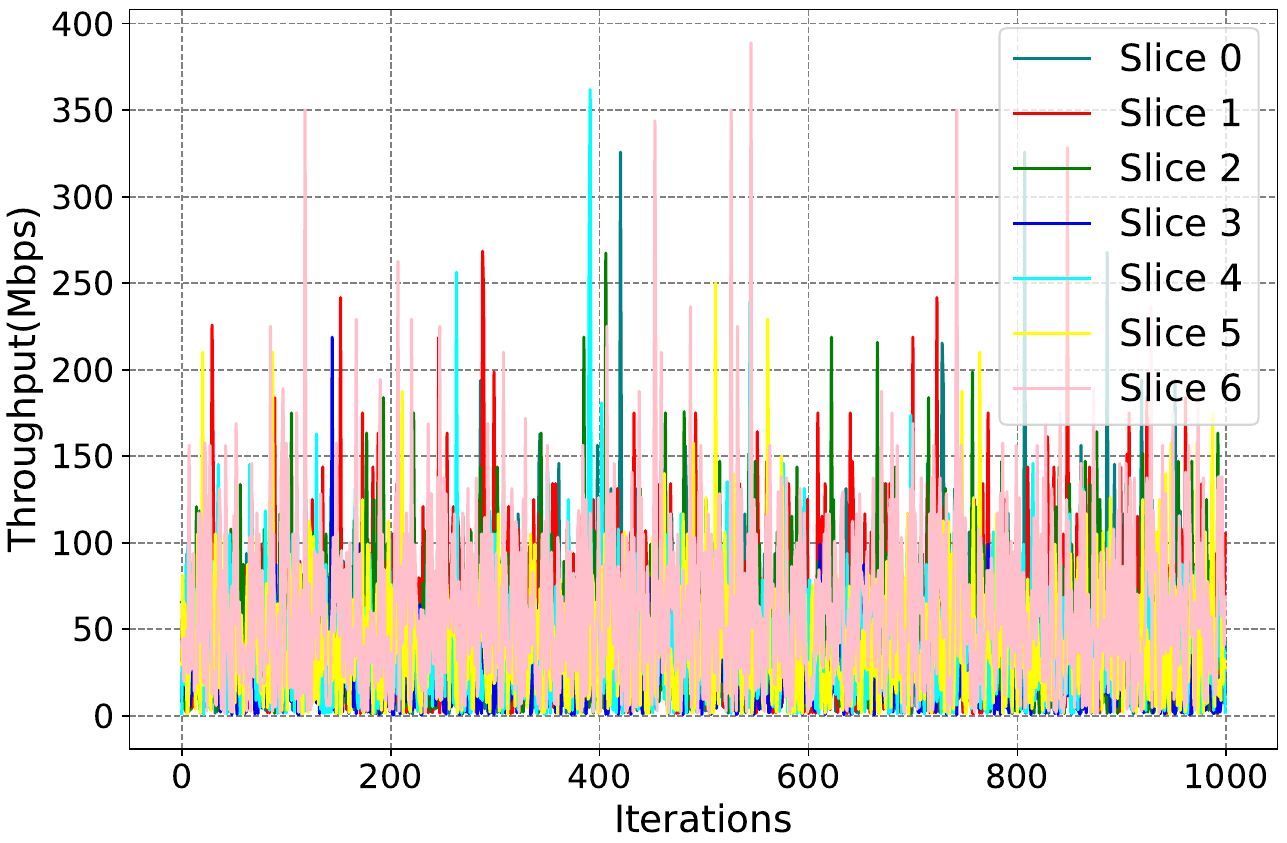}
	\caption{Throughput}
	\label{fig:Throughput}
\end{minipage}
	\begin{minipage}{0.45\textwidth}
		\centering
		\includegraphics[width=1.0\columnwidth]{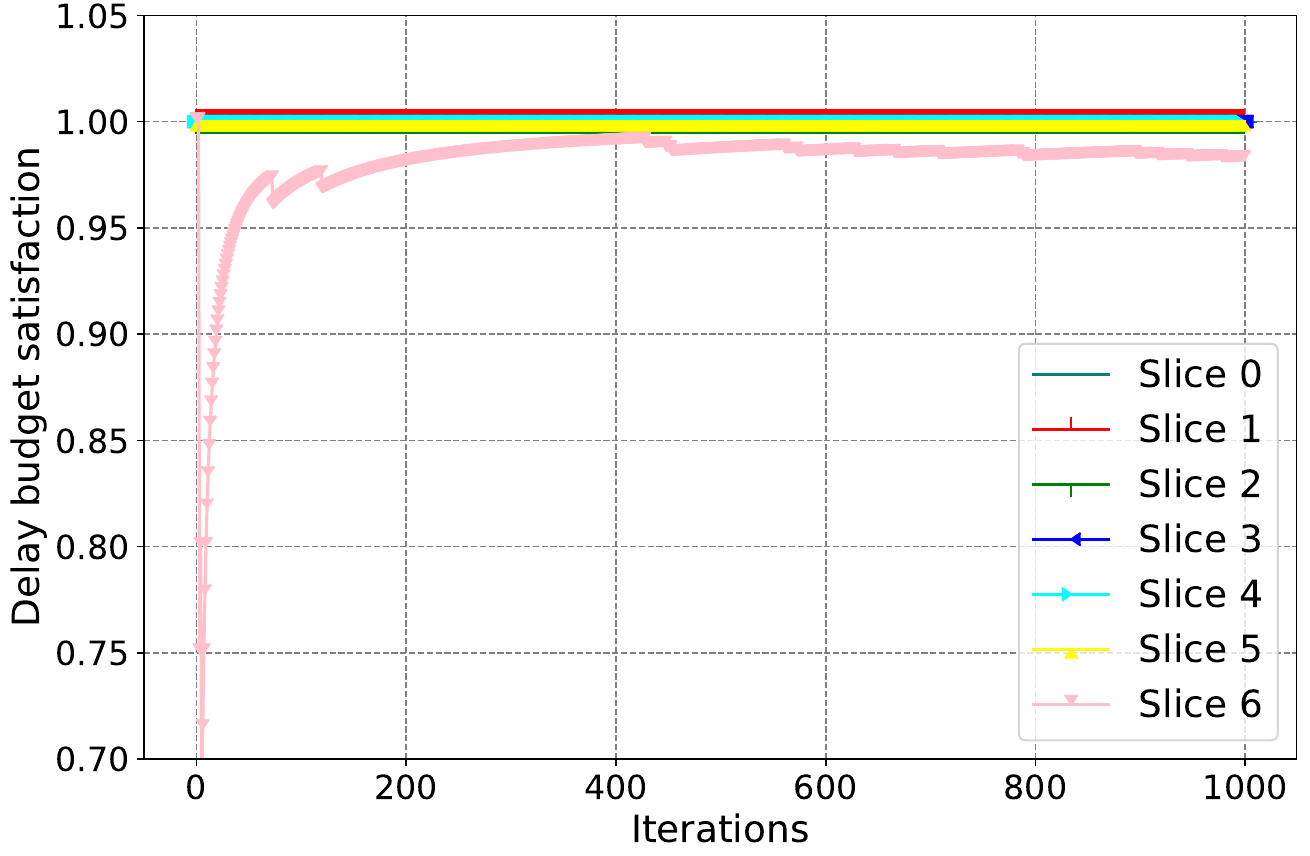}
		\caption{Slice requirement satisfaction (Eq. \ref{eq:network_slice_requirement_satisfaction}) .}
		\label{fig:SliceRequirementSatisfication}
	\end{minipage}
\end{figure}
\begin{figure}[t]
	\begin{minipage}{0.45\textwidth}
		\centering
		\includegraphics[width=1.0\columnwidth]{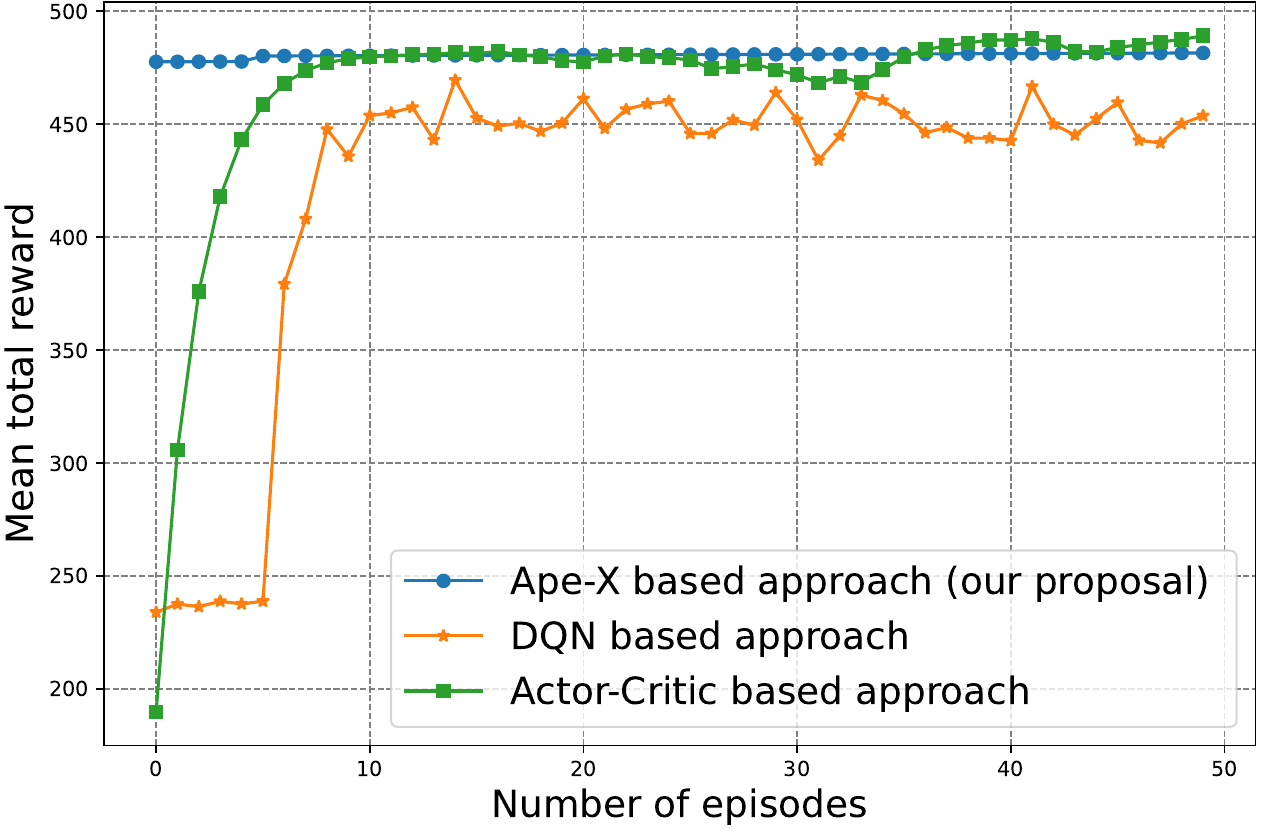}
		\caption{Reward  maximization (Eq. \ref{eq:problem_formulation33}).}
		\label{fig:Apexres}
	\end{minipage}
	\begin{minipage}{0.45\textwidth}
		\centering
		\includegraphics[width=1.0\columnwidth]{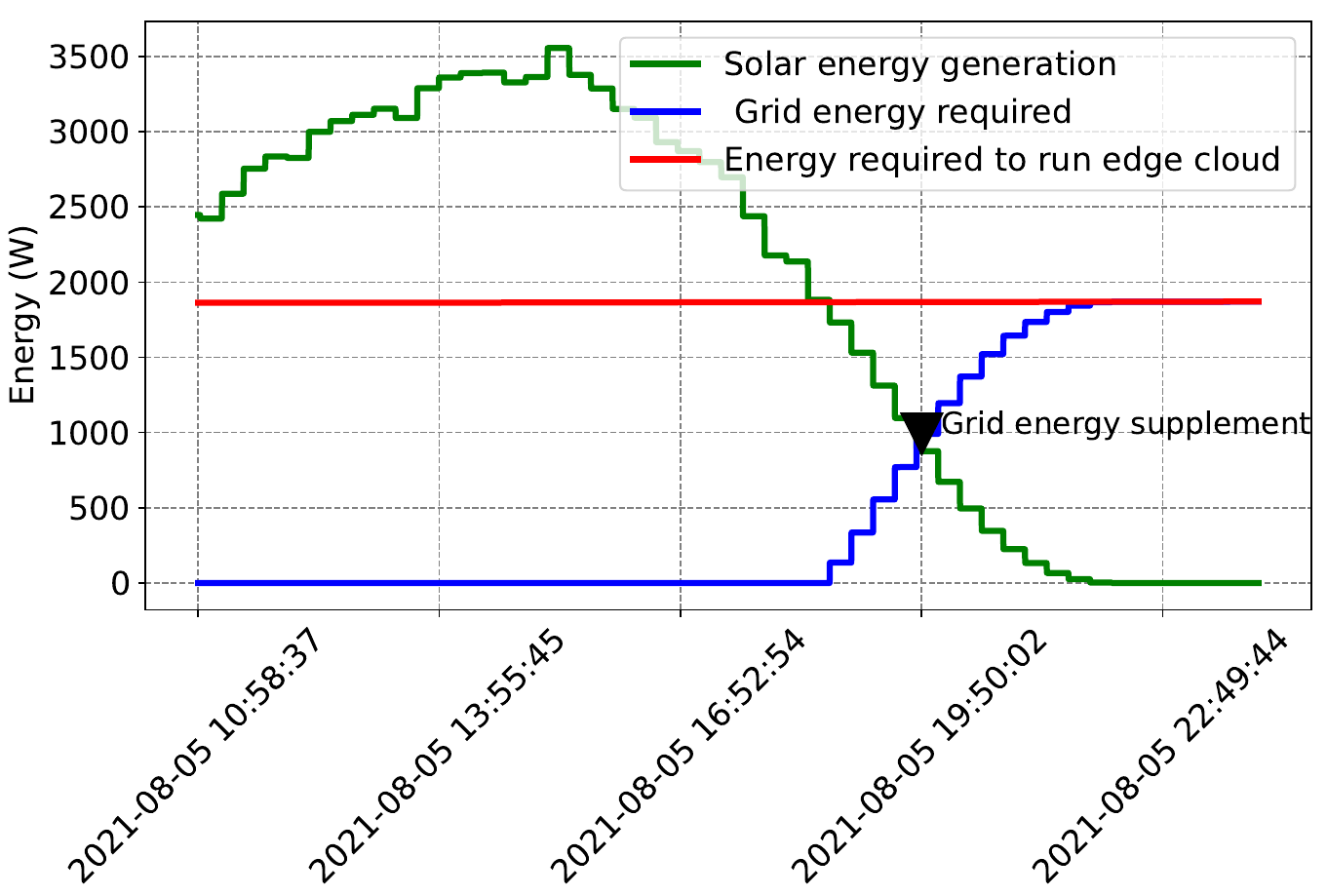}
		\caption{Energy demands.}
		\label{fig:EnergySupplyDemand}
	\end{minipage}
\end{figure}
After prediction, the MEC server sends  $\{\tilde{H}_{T+\tau}\}$  and $\{\tilde{\beta}^{c,k}_{d,{T+\tau}})\}$  to  Near-RT RIC for ultra-small timescale closed-loops and  small-timescale optimization. Here, we consider $\{\tilde{H}_{T+\tau}\}$ and $\{\tilde{\beta}^{c,k}_{d,{T+\tau}})\}$ to be smaller
than $\{H_{1:T}\}$ and $\{\beta^{c,k}_{d,1:T}\}$. Also, the downlink bandwidth is larger than the uplink bandwidth.  Therefore, the time of transmitting $\{\tilde{H}_{T+\tau}\}$ and $\{\tilde{\beta}^{c,k}_{d,{T+\tau}})\}$ is negligible.

\section{Performance Evaluation}
\label{sec:PerformanceEvaluation}

In this section, we evaluate the performance of our proposal. Python is utilized as the programming language for the numerical analysis. 

\subsection{Simulation Setup}
In the performance evaluation for PT,  we consider $6$ O-RUs and $65$ single-family units. CPEs are mounted on single-family units. Since wireless signal strength gradually attenuates as the transmission distance increases, we consider that each CPE  is connected to its nearby O-RU. Furthermore, the services requested at CPEs and corresponding delay budgets are based on the 5QI  \cite{p1} (Table $5.7.4-1$). We choose randomly seven services with 5QI value $1$ for conversational voice ($ \Gamma^k=100$), 5QI value $2$ for conversational video ($ \Gamma^k= 150$), 5QI value $3$ for real-time gaming ($ \Gamma^k= 50$), 5QI value $4$ for non-conversational video ($ \Gamma^k=300$) , 5QI value $7$ for voice and video interactive gaming ($\Gamma^k=100$), 5QI value $70$ for mission-critical data ($ \Gamma^k=200$), and 5QI value  $76$ for live  streaming ($\Gamma^k=500$). The delay budget $\Gamma^k$ is in terms of milliseconds. We use the $28$ GHz band for communication resource, where subcarrier spacing is set to $120$ KHz, numerology $i=3$, and the number of RBs is $ \beta=273$. Fig. \ref{fig:HouseSlice} demonstrates the number of houses per slice. Fig. 	\ref{fig:Slice_vODU} shows the RB per service, where each service is associated with one slice. To schedule RB for CPEs at the edge cloud, we consider three O-DUs and one RT-SC hosted on a single server. These O-DUs are served by two O-CU-UPs, also hosted on the same server. Additionally, the Near-RT RIC, O-CU-CP, and O-CU-UPs are collocated on one server. In total, we have three servers at the edge cloud. Furthermore, at the regional cloud, we consider one server that hosts the Non-RT RIC and UPF, and one MEC server, making a total of two servers in the regional cloud.

For energy consumption for each server at edge cloud, we use the server energy consumption dataset discussed in \cite{9530151} and available in \cite{x6jw-m015-21}, where we have energy consumption from August $2021$ to December $2021$.
Furthermore, for renewable energy, we use historical data of photovoltaic solar power generation in Belgium available in \cite{Elia} for August $2022$. However, solar power generation is for August $2022$, while the server energy consumption dataset is for $2021$. To simplify our evaluation and match the solar power generation dataset with the server energy consumption dataset, we assume that the weather conditions for August 2021 and August $2022$ were almost similar. Since we have a record for every $15$ minutes in the solar energy dataset, to match solar energy with the server energy consumption dataset and RB allocation data, we assume solar power does not change within 15 minutes. Then, we perform data augmentation \cite{bandara2021improving} to have a solar energy record for every second. For the cost of energy and communication resources, we use $\zeta_C  = 6 $ \$ per Mbps, $\tilde{\zeta}_E  = 0.073$ \$ per kWh, and $\zeta_E= 0.070$ \$ per kWh.

We utilize Ray and Keras with TensorFlow \cite{ramasubramanian2019deep} to implement Ape-X in PT and CL in DT. In our Ape-X implementation, the DNN model comprises an input layer with three neurons, two hidden layers with $64$ neurons each, and an output layer with $4$ neurons. The $3$ neurons in the input layer correspond to the $3$ states. In contrast, the $4$ neurons in the output layer represent the $4$ actions: keeping the initial RBs allocation, scaling up RBs, scaling down RBs, and terminating RBs allocation. We assume the initial RBs allocation can be performed based on the auction defined in our previous work \cite{ndikumana2022two}. We set the number of time steps to $100,000$, $\gamma = 0.99$, and the maximum sample size to $50,000$ records.

\subsection{Simulation Results}

Fig. \ref{fig:VDUUsage} demonstrates  RB distribution to O-DUs for scheduling, where O-DU $1$ serves slices $0$, $1$, and $2$, while O-DU $1$ serves slice $3$, $4$, and $5$. O-DU $3$ serves slice $6$. Based on CPE demands and RB scheduled at O-DUs, Fig. 	\ref{fig:VDUUsage} also represents the RB usage ratio at O-DUs, where $1$ corresponds to $100\%$ RB utilization. In other words, above $100\%$ utilization, the demands for RB need to be rejected.
Fig. \ref{fig:Throughput} shows the throughput per each slice, where slice $4$ can reach $350$ $Mbps$.
The data $R^{c,k}_v$ sent to the CPEs was randomly generated from the upper layer, resulting in fluctuations in Figs. \ref{fig:VDUUsage} and  \ref{fig:Throughput}. For future work, we plan to use a dataset specifically for O-RAN based FWA to improve our performance evaluation. Using a real dataset should lead to better performance, as it would not exhibit such fluctuations. Based on RB utilization and delay budget $\Gamma^k$ associated with each service $k$, Fig. 	\ref{fig:SliceRequirementSatisfication} shows the delay budget satisfaction. Here, $1$ corresponds to $100\%$ delay budget satisfaction. The simulation results demonstrate that most slices satisfy the delay budget constraint except slice $6$. The data traffic of slice $6$ has high variations over other slices and this cause also high variations in RB scale-up and scale-own actions. However, the slice $6$ can reach $97\%$ delay budget satisfaction. Furthermore, in maximizing reward (\ref{eq:problem_formulation33}), we compare our  Ape-X based approach with Actor-Critic DRL \cite{rezazadeh2021actor} and Deep Q-Learning \cite{fan2020theoretical}. The simulation results in Fig. \ref{fig:Apexres} demonstrate that our Ape-X based approach performs well than Actor-Critic DRL and Deep Q-Learning in maximizing (\ref{eq:problem_formulation33}).
\begin{figure}[t]
	\centering
	\begin{minipage}{0.45\textwidth}
		\centering
		\includegraphics[width=1.0\columnwidth]{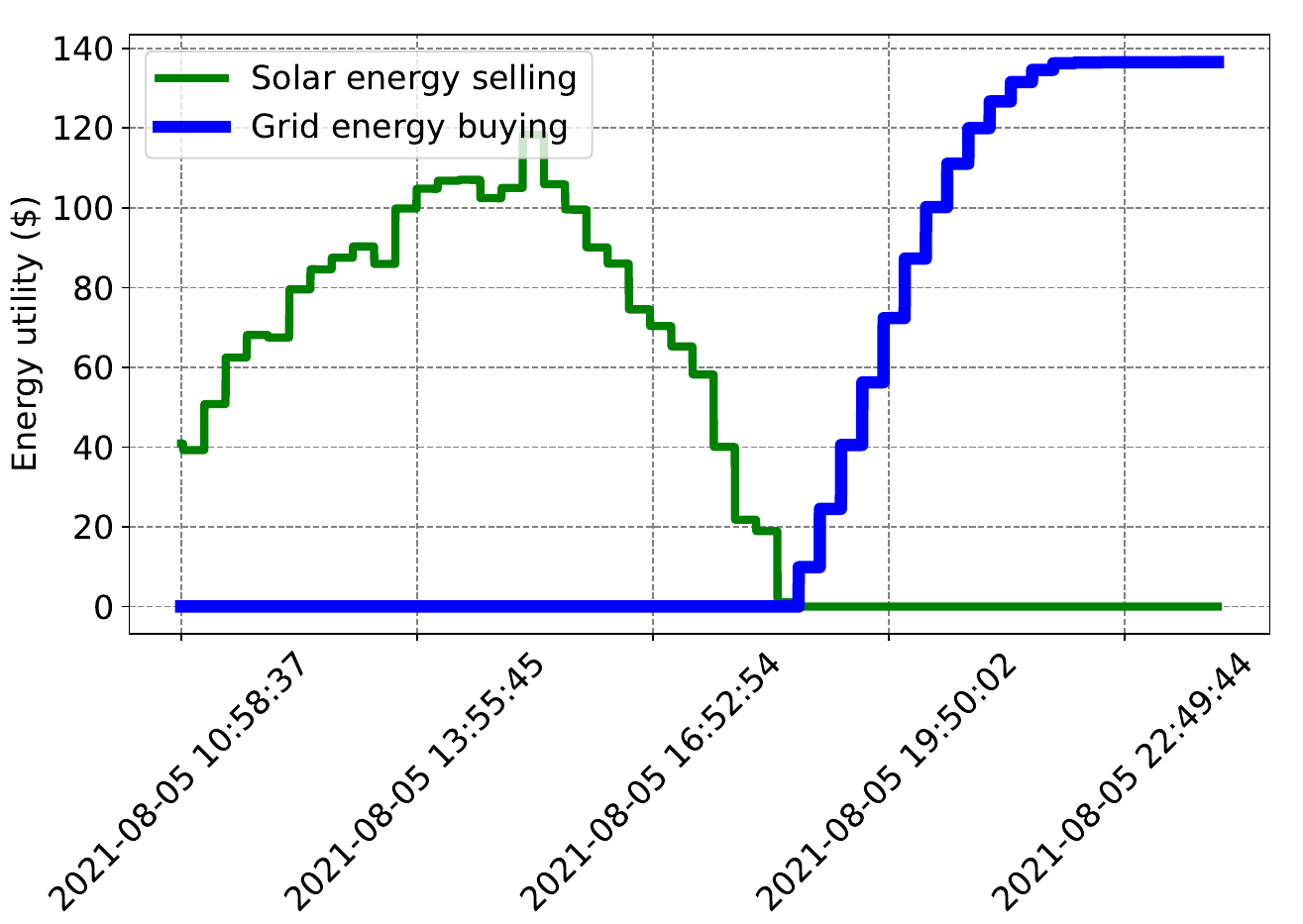}
		\caption{Energy cost.}
		\label{fig:energy_cost}
	\end{minipage}	
\end{figure}
\begin{figure}[t]
		\centering
		\includegraphics[width=1.0\columnwidth]{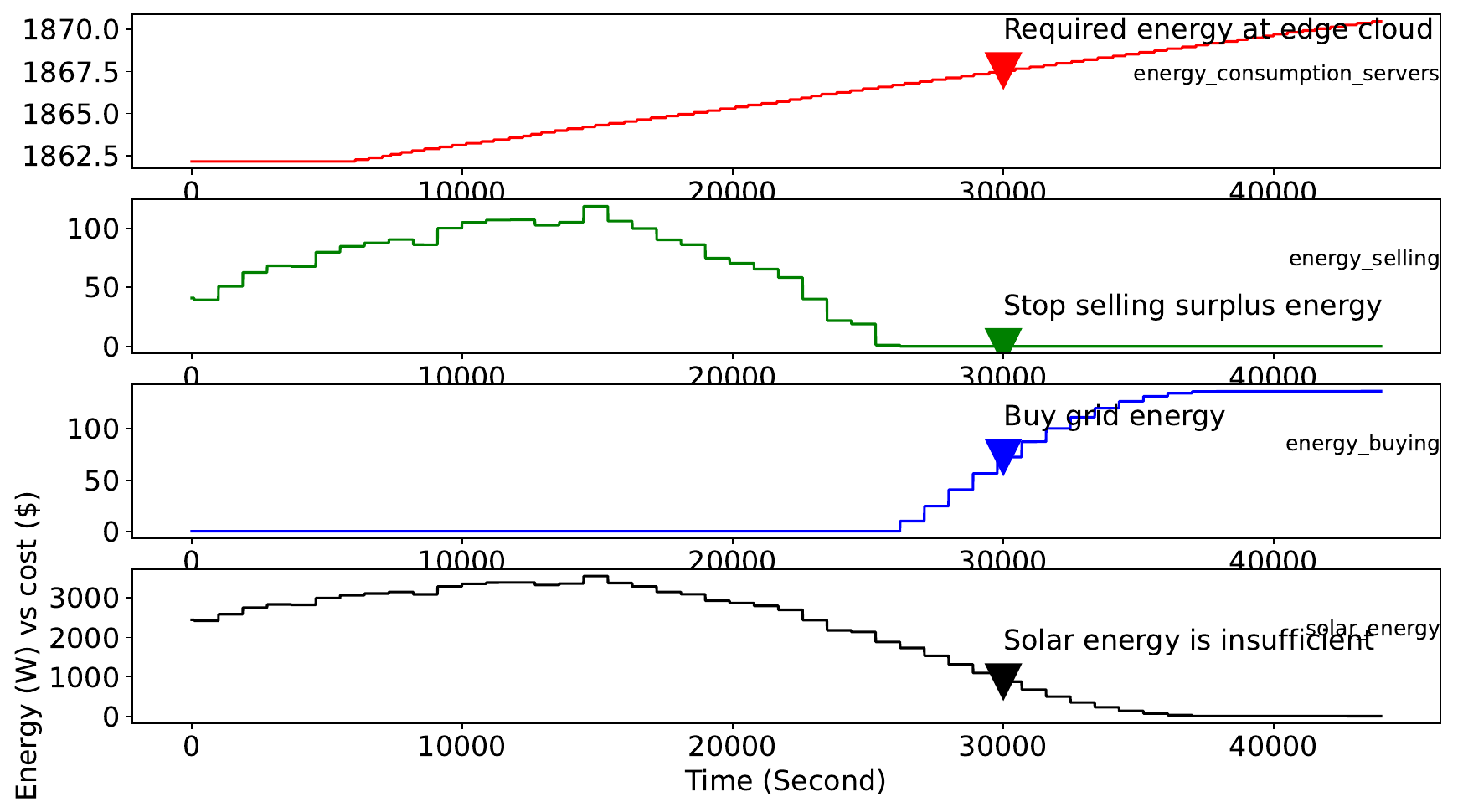}
		\caption{Overview of energy results.}
		\label{fig:EnergySummary}
\end{figure}
\begin{figure}[t]
	\centering
	\begin{minipage}{0.45\textwidth}
		\centering
		\includegraphics[width=1.0\columnwidth]{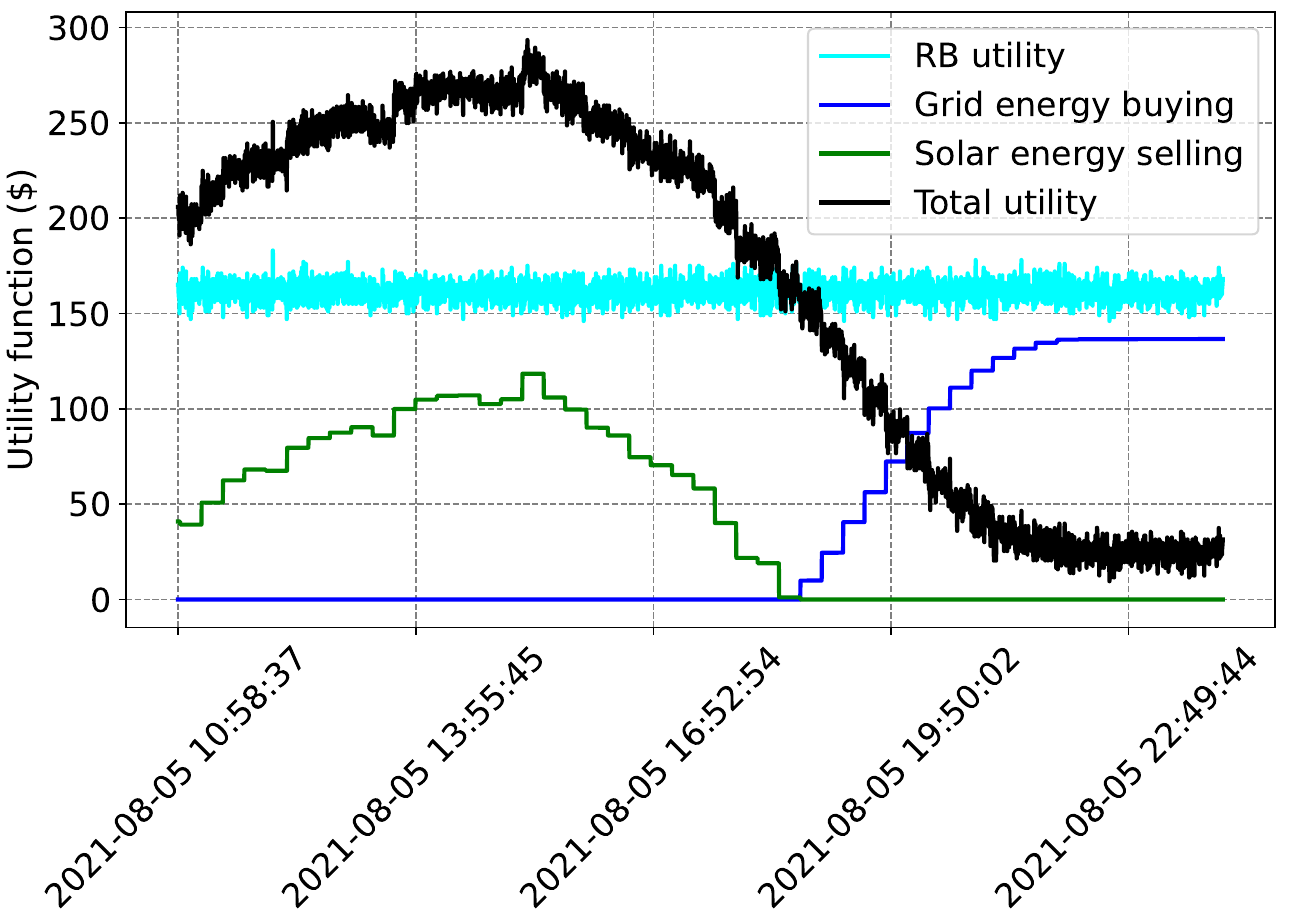}
		\caption{Total utilities and energy cost.}
		\label{fig:Revenue}
	\end{minipage}
		\begin{minipage}{0.45\textwidth}
		\centering
		\includegraphics[width=1.0\columnwidth]{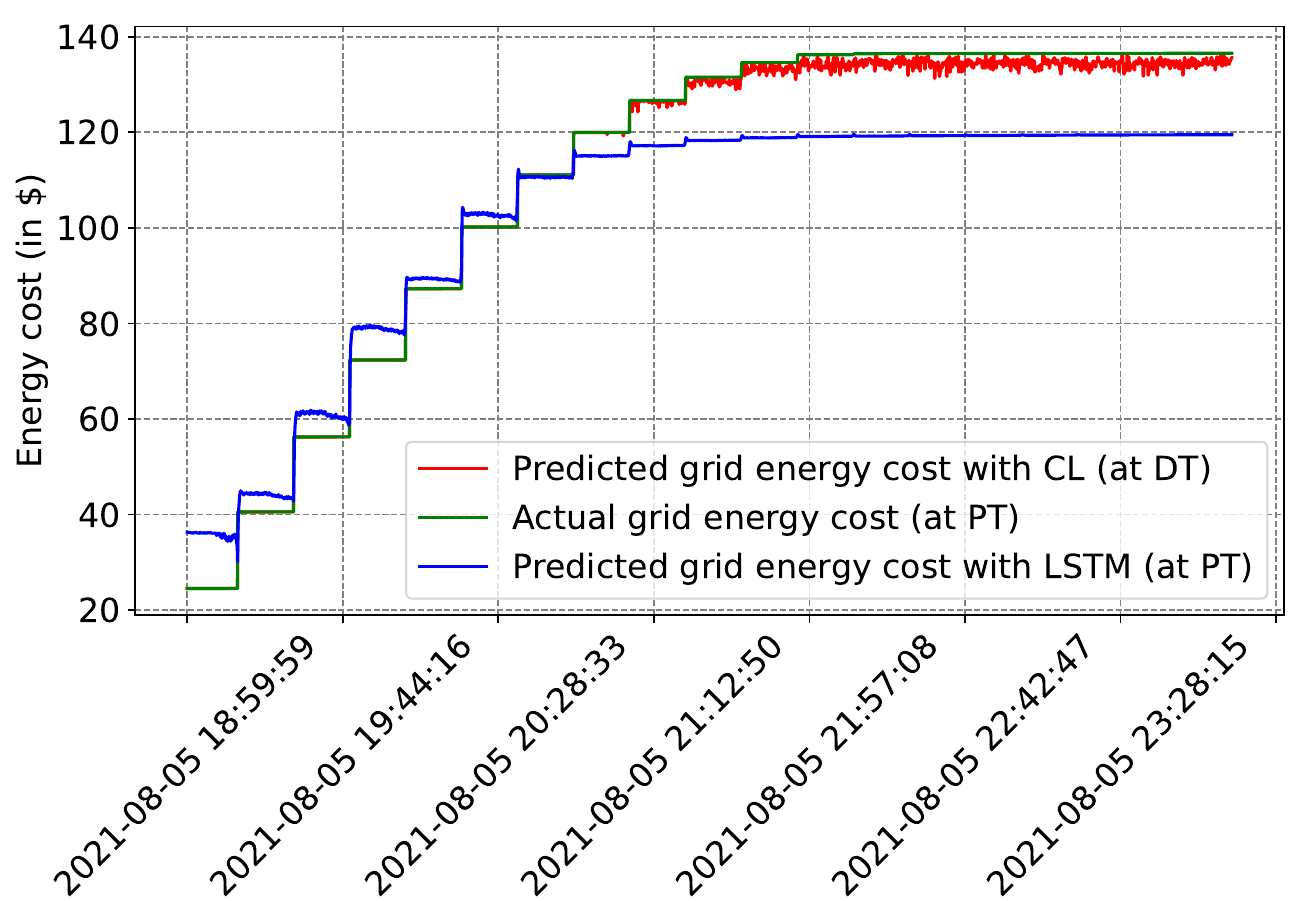}
		\caption{Grid energy cost prediction.}
		\label{fig:Prediction}
	\end{minipage}	
\end{figure}

We consider that the edge cloud is located where the solar energy dataset was generated. Since renewable energy is harvested for free, we can use grid energy when renewable energy is not enough or is exhausted. Fig. \ref{fig:EnergySupplyDemand} shows that we can generate more solar energy from $10$ $AM$ to $5:30$ $PM$, which can satisfy the energy demand of edge cloud. However, from around $7$ $PM$, grid energy has to supplement solar power because solar energy is not enough to meet the energy demand of edge cloud. Fig. \ref{fig:energy_cost} demonstrates that when renewable energy exceeds the required edge cloud energy, the network operator does not buy the grid energy but sells the surplus renewable energy. Here, we remind you that energy storage has limited capacity. In other words, the network operator buys the energy grid when renewable energy is insufficient or is exhausted to run the edge cloud.
Furthermore, the results in Fig. \ref{fig:EnergySummary} show the summary of energy consumption, selling, and buying. 

Figure  \ref{fig:Revenue} depicts the total utility function in black and the communication (i.e., RBs) utility in cyan. During daylight hours, the utility function is significant due to the high availability of solar energy, allowing surplus solar power to be sold to the energy market. At night, the edge cloud provider must allocate funds from the communication utility to purchase grid energy that replaces or supplements the solar energy, consequently reducing the total utility function.

We construct the DT using network topology, RB allocation, and energy data from the PT, which includes 43,966 records. In CL, we use data from $10:58:37$ to $14:58:37$ to train the DNN model and data from $14:58:37$ to $18:59:59$ for testing. The DNN model has a dense input layer with $128$ neurons and a dense output layer with a single neuron. It is trained over $250$ epochs. Once trained and tested, we apply the DNN model within the DT to predict energy costs. Subsequently, we continue collecting every $60$ records from the PT and feed them into the DNN model in the DT to predict the next $60$ records up to $23:59:59$. In other words, in CL, the DNN model at the DT adapts progressively as new PT data $\{H_{1:T}\} $ and $\{\beta^{c,k}_{d,1:T}\}$ arrives (in sets of $60$ records) to reduce a large gap between the DT and the PT. We then compare the CL predictions with those generated by a Long Short-Term Memory (LSTM) model. The LSTM model has an input layer of $64$ neurons, two hidden layers of $64$ neurons each, and an output layer with one neuron. For the LSTM, we use $26,379$ records for training, $17,587$ for testing, and a lookback window of $7,200$ records. The LSTM is trained and tested once and then used for prediction. In our case, the LSTM can be implemented either in the PT or DT because it does not require continuous learning. On the other hand, CL performs better when running in DT because it adapts progressively as new PT data arrives. In Fig.  \ref{fig:Prediction}, the simulation results show that the CL in DT outperforms the LSTM in predicting energy cost. The similar predictive approach can be applied to predict communication (i.e., RBs) utility.

\section{Conclusion}
\label{sec:Conclusion}
In this work, we presented O-RAN closed-loops for energy-efficient slice resource management in 5G FWA PT assisted by the DT model. We utilized one closed-loop to distribute radio resources to O-RAN elements hosted at the edge cloud and to manage slices for scheduling. We designed another closed-loop for intra-slice resource allocation for CPEs. We introduced an energy model and integrated it with radio resource allocation. We then formulated ultra-small and small-timescale optimizations for PT to maximize RB utilization and delay budget satisfaction while minimizing energy costs. We transformed the optimization problems into reward functions and devised a reinforcement learning technique to maximize these functions. Subsequently, we applied the majorization-minimization technique to solve the jointly formulated problem for the energy and communication models. Finally, we presented a DT that replicates PT by incorporating solution experiences into future states. The numerical results from August demonstrate that the proposed approach can efficiently utilize energy resources while maximizing slice utilization and satisfying delay budget requirements. One of our future goals is to expand our performance evaluation to include large-scale energy-efficient slice resource management in 5G O-RAN based FWA.
\bibliographystyle{IEEEtran}

\begin{thebibliography}{10}
\providecommand{\url}[1]{#1}
\csname url@samestyle\endcsname
\providecommand{\newblock}{\relax}
\providecommand{\bibinfo}[2]{#2}
\providecommand{\BIBentrySTDinterwordspacing}{\spaceskip=0pt\relax}
\providecommand{\BIBentryALTinterwordstretchfactor}{4}
\providecommand{\BIBentryALTinterwordspacing}{\spaceskip=\fontdimen2\font plus
\BIBentryALTinterwordstretchfactor\fontdimen3\font minus
  \fontdimen4\font\relax}
\providecommand{\BIBforeignlanguage}[2]{{%
\expandafter\ifx\csname l@#1\endcsname\relax
\typeout{** WARNING: IEEEtran.bst: No hyphenation pattern has been}%
\typeout{** loaded for the language `#1'. Using the pattern for}%
\typeout{** the default language instead.}%
\else
\language=\csname l@#1\endcsname
\fi
#2}}
\providecommand{\BIBdecl}{\relax}
\BIBdecl

\bibitem{adityo20215g}
M.~K. Adityo, M.~I. Nashiruddin, and M.~A. Nugraha, ``{5G} fixed wireless
  access network for urban residential market: A case of indonesia,'' in
  \emph{Proceedings of IEEE International Conference on Internet of Things and
  Intelligence Systems ({IoTaIS})}.\hskip 1em plus 0.5em minus 0.4em\relax
  IEEE, 2021, pp. 123--128.

\bibitem{laraqui2017fixed}
K.~Laraqui, S.~Tombaz, A.~Furusk{\"a}r, B.~Skubic, A.~Nazari, and E.~Trojer,
  ``Fixed wireless access: On a massive scale with {5G},'' \emph{Ericsson
  review (English ed.)}, vol.~94, no.~1, pp. 52--65, 2017.

\bibitem{5gamericas}
G.~americas, ``Fixed wireless access with {5G} networks,'' \emph{{5G} Americas
  White Paper}, November, 2021.

\bibitem{allianceORANUseCases}
O.~Alliance, ``{O-RAN} work group 1 (use cases and overall architecture),
  {O-RAN} architecture description,'' \emph{O-RAN.WG1.OAD-R003-v12.00, June},
  2024.

\bibitem{isik2023architectural}
G.~E. ISIK and H.~ACHTEN, ``Architectural hybrid (physical-digital) prototyping
  in design processes with digital twin technologies,'' \emph{Architecture and
  Planning Journal (APJ)}, vol.~28, no.~3, p.~4, 2023.

\bibitem{mihai2022digital}
S.~Mihai, M.~Yaqoob, D.~V. Hung, W.~Davis, P.~Towakel, M.~Raza, M.~Karamanoglu,
  B.~Barn, D.~Shetve, R.~V. Prasad \emph{et~al.}, ``Digital twins: a survey on
  enabling technologies, challenges, trends and future prospects,'' \emph{IEEE
  Communications Surveys \& Tutorials}, 2022.

\bibitem{haag2018digital}
S.~Haag and R.~Anderl, ``Digital twin--proof of concept,'' \emph{Manufacturing
  letters}, vol.~15, pp. 64--66, 2018.

\bibitem{martiradonna2021cascaded}
S.~Martiradonna, G.~Cisotto, G.~Boggia, G.~Piro, L.~Vangelista, and S.~Tomasin,
  ``Cascaded wlan-fwa networking and computing architecture for pervasive
  in-home healthcare,'' \emph{IEEE Wireless Communications}, vol.~28, no.~3,
  pp. 92--99, 2021.

\bibitem{chuah2020intelligent}
T.~C. Chuah and Y.~L. Lee, ``Intelligent ran slicing for broadband access in
  the 5g and big data era,'' \emph{IEEE Communications Magazine}, vol.~58,
  no.~8, pp. 69--75, 2020.

\bibitem{boutaba2021ai}
R.~Boutaba, N.~Shahriar, M.~A. Salahuddin, S.~R. Chowdhury, N.~Saha, and
  A.~James, ``Ai-driven closed-loop automation in {5G} and beyond mobile
  networks,'' in \emph{Proceedings of the 4th FlexNets Workshop on Flexible
  Networks Artificial Intelligence Supported Network Flexibility and Agility},
  2021, pp. 1--6.

\bibitem{rahmawati2022assessing}
P.~Rahmawati, M.~I. Nashiruddin, A.~T. Hanuranto, and A.~Akhmad, ``Assessing
  3.5 ghz frequency for {5G} new radio (nr) implementation in indonesia's urban
  area,'' in \emph{Proceedings of 12th Annual Computing and Communication
  Workshop and Conference (CCWC)}.\hskip 1em plus 0.5em minus 0.4em\relax IEEE,
  2022, pp. 0876--0882.

\bibitem{lappalainen2021planning}
A.~Lappalainen, Y.~Zhang, and C.~Rosenberg, ``Planning {5G} networks for rural
  fixed wireless access,'' \emph{arXiv preprint arXiv:2110.01456}, 2021.

\bibitem{van2019three}
G.~M. Van~de Ven and A.~S. Tolias, ``Three scenarios for continual learning,''
  \emph{arXiv preprint arXiv:1904.07734}, 2019.

\bibitem{hashemi2017integrated}
M.~Hashemi, M.~Coldrey, M.~Johansson, and S.~Petersson, ``Integrated access and
  backhaul in fixed wireless access systems,'' in \emph{Proceedings of 86th
  Vehicular Technology Conference (VTC-Fall)}.\hskip 1em plus 0.5em minus
  0.4em\relax IEEE, 2017, pp. 1--5.

\bibitem{matrakidis2021converged}
C.~Matrakidis, E.~Kosmatos, A.~Stavdas, P.~Kostopoulos, D.~Uzunidis,
  S.~Horlitz, T.~Pfeiffer, and A.~Lord, ``A converged fixed-wireless tdma-based
  infrastructure exploiting qos-aware end-to-end slicing,'' in
  \emph{Proceedings of European Conference on Optical Communication
  (ECOC)}.\hskip 1em plus 0.5em minus 0.4em\relax IEEE, 2021, pp. 1--4.

\bibitem{larsen2021energy}
L.~M. Larsen, S.~Ruepp, M.~S. Berger, and H.~L. Christiansen, ``Energy
  consumption modelling of next generation mobile crosshaul networks,'' in
  \emph{Proceedings of International Conferences on Internet of Things
  (iThings) and IEEE Green Computing \& Communications (GreenCom) and IEEE
  Cyber, Physical \& Social Computing (CPSCom) and IEEE Smart Data (SmartData)
  and IEEE Congress on Cybermatics (Cybermatics)}.\hskip 1em plus 0.5em minus
  0.4em\relax IEEE, 2021, pp. 153--160.

\bibitem{kaliski2023supporting}
R.~Kaliski, S.-M. Cheng, and C.-F. Hung, ``Supporting {6G} mission-critical
  services on {O-RAN},'' \emph{Proceedings of Internet of Things Magazine},
  vol.~6, no.~3, pp. 32--37, 2023.

\bibitem{miyazawa2024energy}
T.~Miyazawa, K.~Ishizu, H.~Asaeda, H.~Tsuji, and H.~Harai, ``Energy-efficient
  power management for o-ran base stations utilizing pedestrian flow analytics
  and non-terrestrial networks,'' \emph{IEICE Transactions on Communications},
  2024.

\bibitem{dinh2020home}
H.~T. Dinh, J.~Yun, D.~M. Kim, K.-H. Lee, and D.~Kim, ``A home energy
  management system with renewable energy and energy storage utilizing main
  grid and electricity selling,'' \emph{IEEE Access}, vol.~8, pp.
  49\,436--49\,450, 2020.

\bibitem{azimi2021energy}
Y.~Azimi, S.~Yousefi, H.~Kalbkhani, and T.~Kunz, ``Energy-efficient deep
  reinforcement learning assisted resource allocation for {5G-RAN} slicing,''
  \emph{IEEE Transactions on Vehicular Technology}, vol.~71, no.~1, pp.
  856--871, 2021.

\bibitem{pamuklu2021energy}
T.~Pamuklu, S.~Mollahasani, and M.~Erol-Kantarci, ``Energy-efficient and
  delay-guaranteed joint resource allocation and du selection in o-ran,'' in
  \emph{Proceedings of 4th {5G} World Forum (5GWF)}.\hskip 1em plus 0.5em minus
  0.4em\relax IEEE, 2021, pp. 99--104.

\bibitem{chang2018radio}
C.-Y. Chang, N.~Nikaein, and T.~Spyropoulos, ``Radio access network resource
  slicing for flexible service execution,'' in \emph{Proceedings of IEEE
  Conference on Computer Communications Workshops (INFOCOM WKSHPS)}.\hskip 1em
  plus 0.5em minus 0.4em\relax IEEE, 2018, pp. 668--673.

\bibitem{xie2019towards}
M.~Xie, W.~Y. Poe, Y.~Wang, A.~J. Gonzalez, A.~M. Elmokashfi, J.~A.~P.
  Rodrigues, and F.~Michelinakis, ``Towards closed loop {5G} service assurance
  architecture for network slices as a service,'' in \emph{Proceedings of
  European Conference on Networks and Communications (EuCNC)}.\hskip 1em plus
  0.5em minus 0.4em\relax IEEE, 2019, pp. 139--143.

\bibitem{naik2022closed}
P.~Naik, C.~Govindarajan, S.~Goel, K.~Govindarajan, D.~Behl, A.~Singh,
  M.~Thomas, U.~Mangla, and P.~Jayachandran, ``Closed-loop automation for {5G}
  slice assurance,'' in \emph{Proceedings of 14th International Conference on
  COMmunication Systems \& NETworkS (COMSNETS)}.\hskip 1em plus 0.5em minus
  0.4em\relax IEEE, 2022, pp. 424--426.

\bibitem{9833928}
O.~Hashash, C.~Chaccour, and W.~Saad, ``Edge continual learning for dynamic
  digital twins over wireless networks,'' in \emph{Proceedings of 23rd
  International Workshop on Signal Processing Advances in Wireless
  Communication (SPAWC)}.\hskip 1em plus 0.5em minus 0.4em\relax IEEE, 2022,
  pp. 1--5.

\bibitem{peng2022distributed}
K.~Peng, H.~Huang, M.~Bilal, and X.~Xu, ``Distributed incentives for
  intelligent offloading and resource allocation in digital twin driven smart
  industry,'' \emph{IEEE Transactions on Industrial Informatics}, 2022.

\bibitem{10437288}
A.~Ndikumana, K.~K. Nguyen, and M.~Cheriet, ``Digital twin assisted
  closed-loops for energy-efficient open ran-based fixed wireless access
  provisioning in rural areas,'' in \emph{Proceedings of GLOBECOM 2023 - 2023
  IEEE Global Communications Conference}, 2023, pp. 6285--6290.

\bibitem{saba2024using}
N.~Saba, J.~Salo, K.~Ruttik, and R.~J{\"a}ntti, ``Using existing base station
  sites for 5g millimeter-wave fixed wireless access: Antenna height and
  coverage analysis,'' in \emph{Proceedings of IEEE Wireless Communications and
  Networking Conference (WCNC)}.\hskip 1em plus 0.5em minus 0.4em\relax IEEE,
  2024, pp. 1--6.

\bibitem{ndikumana2019joint}
A.~Ndikumana, N.~H. Tran, T.~M. Ho, Z.~Han, W.~Saad, D.~Niyato, and C.~S. Hong,
  ``Joint communication, computation, caching, and control in big data
  multi-access edge computing,'' \emph{IEEE Transactions on Mobile Computing},
  vol.~19, no.~6, pp. 1359--1374, 2019.

\bibitem{ndikumana2022age}
A.~Ndikumana, K.~K. Nguyen, and M.~Cheriet, ``Age of processing-based data
  offloading for autonomous vehicles in {MultiRATs} open {RAN},'' \emph{IEEE
  Transactions on Intelligent Transportation Systems}, 2022.

\bibitem{deng2013multigreen}
W.~Deng, F.~Liu, H.~Jin, C.~Wu, and X.~Liu, ``Multigreen: Cost-minimizing
  multi-source datacenter power supply with online control,'' in
  \emph{Proceedings of the fourth international conference on Future energy
  systems}, 2013, pp. 149--160.

\bibitem{kanhere2002fair}
S.~S. Kanhere, H.~Sethu, and A.~B. Parekh, ``Fair and efficient packet
  scheduling using elastic round robin,'' \emph{IEEE transactions on Parallel
  and Distributed Systems}, vol.~13, no.~3, pp. 324--336, 2002.

\bibitem{ndikumana2022two}
A.~Ndikumana, N.~Kim~Khoa, and M.~Cheriet, ``Two-level closed loops for {RAN}
  slice resources management serving flying and ground-based cars,'' \emph{IEEE
  Transactions on Network and Service Management}, 2023.

\bibitem{138211}
{ETSI-TS}, ``{5G} {NR} physical channels and modulation (3gpp ts 38.211 version
  17.2.0 release 17) ({ETSI} {TS} 138 211 v17.2.0),'' 2022-07.

\bibitem{etsi1308}
{ETSI}, ``User equipment ({UE}) radio access capabilities ({3GPP} {TS} 38.306
  version 17.0.0 release 17),'' 2022-05.

\bibitem{yoo2004capacity}
T.~Yoo and A.~Goldsmith, ``Capacity of fading mimo channels with channel
  estimation error,'' in \emph{Proceedings of IEEE International Conference on
  Communications (IEEE Cat. No. 04CH37577)}, vol.~2.\hskip 1em plus 0.5em minus
  0.4em\relax IEEE, 2004, pp. 808--813.

\bibitem{petri2021edge}
I.~Petri, O.~Rana, Y.~Rezgui, and F.~Fadli, ``Edge hvac analytics,''
  \emph{Energies}, vol.~14, no.~17, p. 5464, 2021.

\bibitem{kiran2021deep}
B.~R. Kiran, I.~Sobh, V.~Talpaert, P.~Mannion, A.~A. Al~Sallab, S.~Yogamani,
  and P.~P{\'e}rez, ``Deep reinforcement learning for autonomous driving: A
  survey,'' \emph{IEEE Transactions on Intelligent Transportation Systems},
  2021.

\bibitem{horgan2018distributed}
D.~Horgan, J.~Quan, D.~Budden, G.~Barth-Maron, M.~Hessel, H.~Van~Hasselt, and
  D.~Silver, ``Distributed prioritized experience replay,'' \emph{arXiv
  preprint arXiv:1803.00933}, 2018.

\bibitem{scutari2018parallel}
G.~Scutari and Y.~Sun, ``Parallel and distributed successive convex
  approximation methods for big-data optimization,'' in \emph{Multi-agent
  Optimization}.\hskip 1em plus 0.5em minus 0.4em\relax Springer, 2018, pp.
  141--308.

\bibitem{stellato2020osqp}
B.~Stellato, G.~Banjac, P.~Goulart, A.~Bemporad, and S.~Boyd, ``Osqp: An
  operator splitting solver for quadratic programs,'' \emph{Mathematical
  Programming Computation}, vol.~12, no.~4, pp. 637--672, 2020.

\bibitem{schuster2020exact}
R.~Schuster, J.~O. Hanson, M.~Strimas-Mackey, and J.~R. Bennett, ``Exact
  integer linear programming solvers outperform simulated annealing for solving
  conservation planning problems,'' \emph{PeerJ}, vol.~8, p. e9258, 2020.

\bibitem{p1}
3GPP, ``{3GPP} {TS} 23.501 (2019), system architecture for the {5G} system;
  stage 2.'' \emph{3GPP TS 23.501}, 2019.

\bibitem{9530151}
V.~Asanza, R.~E. Pico, D.~Torres, S.~Santillan, and J.~Cadena, ``{FPGA} based
  meteorological monitoring station,'' in \emph{Proceedings of 2021 IEEE
  Sensors Applications Symposium (SAS)}, 2021, pp. 1--6.

\bibitem{x6jw-m015-21}
\BIBentryALTinterwordspacing
A.~Bazurto, D.~Torres, V.~Asanza, and R.~Estrada, ``Data server energy
  consumption dataset,'' 2021. [Online]. Available:
  \url{https://dx.doi.org/10.21227/x6jw-m015}
\BIBentrySTDinterwordspacing

\bibitem{Elia}
{Elia Group}, ``Solar power generation for belgium,''
  \url{https://www.elia.be/en/grid-data/power-generation/solar-pv-power-generation-data},
  [Online; accessed October. 20, 2022].

\bibitem{bandara2021improving}
K.~Bandara, H.~Hewamalage, Y.-H. Liu, Y.~Kang, and C.~Bergmeir, ``Improving the
  accuracy of global forecasting models using time series data augmentation,''
  \emph{Pattern Recognition}, vol. 120, p. 108148, 2021.

\bibitem{ramasubramanian2019deep}
K.~Ramasubramanian and A.~Singh, ``Deep learning using keras and tensorflow,''
  in \emph{Machine Learning Using R}.\hskip 1em plus 0.5em minus 0.4em\relax
  Springer, 2019, pp. 667--688.

\bibitem{rezazadeh2021actor}
F.~Rezazadeh, H.~Chergui, L.~Christofi, and C.~Verikoukis, ``Actor-critic-based
  learning for zero-touch joint resource and energy control in network
  slicing,'' in \emph{Proceedings of IEEE International Conference on
  Communications (ICC)}.\hskip 1em plus 0.5em minus 0.4em\relax IEEE, 2021, pp.
  1--6.

\bibitem{fan2020theoretical}
J.~Fan, Z.~Wang, Y.~Xie, and Z.~Yang, ``A theoretical analysis of deep
  q-learning,'' in \emph{Proceedings of Learning for Dynamics and
  Control}.\hskip 1em plus 0.5em minus 0.4em\relax PMLR, 2020, pp. 486--489.

\end{thebibliography}

\begin{IEEEbiography}[{\includegraphics[width=1in,height=1.25in,clip,keepaspectratio]{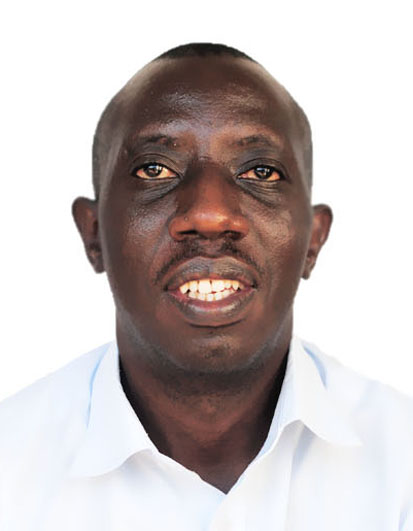}}]{Anselme Ndikumana} received  B.S. degree in Computer Science from the National University of Rwanda in 2007 and Ph.D. degree in Computer Engineering from Kyung Hee University, South Korea in August 2019. Since 2020, he has been	with the Synchromedia Lab, École de Technologie Supérieure, Université du Québec, Montréal, QC, Canada, where he is currently a Postdoctoral Researcher. His professional experience includes Lecturer at the University of Lay Adventists of Kigali from 2019 to 2020, Chief Information System, a System Analyst, and a Database Administrator at Rwanda Utilities Regulatory Authority from 2008 to 2014. His research interest includes AI for wireless communication, multi-access edge computing, 5G and next-G networks, metaverse, network economics, game theory,  and network optimization.
\end{IEEEbiography}
\begin{IEEEbiography}[{\includegraphics[width=1in,height=1.25in,clip,keepaspectratio]{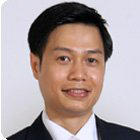}}]{Kim Khoa Nguyen} is Associate Professor in the Department of Electrical Engineering and the Founder and Director of the IoT and Cloud Computing Laboratory at the University of Quebec’s Ecole de technologie supérieure, Montreal, Canada. He holds the VMware-Broadcom Industrial Chair in Multi-Cloud Service Grid and Edge AI. In the past, he served as CTO of Inocybe Technologies (now is Kontron Canada), a world’s leading company in software-defined networking (SDN) solutions. He led R\&D in several large-scale projects with world-class corporations such as VMware, Ericsson, Ciena, Telus, InterDigital, and Ultra Electronics. He is the recipient of the Microsoft’s Azure Global IoT Contest Award 2017, the Ciena’s Aspirational Prize 2018, and the IEEE Future Internet’s “Connecting the Unconnected” Award 2023. He is the author of more than 180 publications and holds several industrial patents. His expertise includes network optimization, cloud computing, IoT, 5G, machine learning, AI, smart city, high speed networks, and green ICT. 
\end{IEEEbiography}
\begin{IEEEbiography}[{\includegraphics[width=1in,height=1.25in,clip,keepaspectratio]{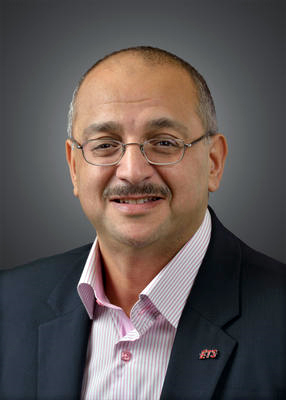}}]{Dr. Mohamed Cheriet} received his Bachelor, M.Sc. and Ph.D. degrees in Computer Science from USTHB (Algiers) and the University of Pierre \& Marie Curie (Paris VI) in 1984, 1985 and 1988 respectively. He was then a Postdoctoral Fellow at CNRS, Pont et Chaussées, Paris V, in 1988, and at CENPARMI, Concordia U., Montreal, in 1990. Since 1992, he has been a professor in the Systems Engineering department at the University of Quebec - École de Technologie Supérieure (ÉTS), Montreal, and was appointed full Professor there in 1998. Prof. Cheriet was the director of LIVIA Laboratory for Imagery, Vision, and Artificial Intelligence (2000-2006), and is the founder and director of Synchromedia Laboratory for multimedia communication in telepresence applications, since 1998.  Dr. Cheriet research has extensive experience in Sustainable and Intelligent Next Generation Systems. Dr. Cheriet is an expert in Computational Intelligence, Pattern Recognition, Machine Learning, Artificial Intelligence and Perception and their applications, more extensively in Networking and Image Processing. In addition, Dr. Cheriet has published more than 500 technical papers in the field and serves on the editorial boards of several renowned journals and international conferences. He held a Tier 1 Canada Research Chair on Sustainable and Smart Eco-Cloud (2013-2000), and lead the establishment of the first smart university campus in Canada, created as a hub for innovation and productivity at Montreal. Dr. Cheriet is the General Director of the FRQNT Strategic Cluster on the Operationalization of Sustainability Development, CIRODD (2019-2026). He is the Administrative Director of the \$12M CFI’2022 CEOS*Net Manufacturing Cloud Network. He is a 2016 Fellow of the International Association of Pattern Recognition (IAPR), a 2017 Fellow of the Canadian Academy of Engineering (CAE), a 2018 Fellow of the Engineering Institute of Canada (EIC), and a 2019 Fellow of Engineers Canada (EC). Dr. Cheriet is the recipient of the 2016 IEEE J.M. Ham Outstanding Engineering Educator Award, the 2013 ÉTS Research Excellence prize, for his outstanding contribution in green ICT, cloud computing, and big data analytics research areas, and the 2012 Queen Elizabeth II Diamond Jubilee Medal. He is a senior member of the IEEE, the founder and former Chair of the IEEE Montreal Chapter of Computational Intelligent Systems (CIS), a Steering Committee Member of the IEEE Sustainable ICT Initiative, and the Chair of ICT Emissions Working Group. He contributed 6 patents (3 granted), and the first standard ever, IEEE 1922.2, on real-time calculation of ICT emissions, in April 2020, with his IEEE Emissions Working Group.
\end{IEEEbiography}
\end{document}